\documentclass[journal]{IEEEtran}
\ifCLASSINFOpdf
\else
\fi

\usepackage{graphicx}
\usepackage{amssymb}
\usepackage{amsmath}
\usepackage{booktabs}
\usepackage{multirow}

\graphicspath{{figures/}}

\usepackage{xcolor}
\definecolor{ECS-Blue}{rgb}{0,0.4392,0.7529}
\definecolor{ECS-Red}{rgb}{0.7529,0,0}
\definecolor{ECS-Green}{rgb}{0.4667,0.5765,0.2353}

\usepackage[para]{threeparttable}
\makeatletter
\def\TPT@doparanotes{\par
   \prevdepth\z@ \TPT@hsize
   \TPTnoteSettings
   \parindent\z@ \pretolerance 8
   \linepenalty 200
   \renewcommand\item[1][]{\relax\ifhmode \begingroup
       \unskip
       \advance\hsize 10em 
       \penalty -45 \hskip\z@\@plus\hsize \penalty-19
       \hskip .15\hsize \penalty 9999 \hskip-.15\hsize
       \hskip .01\hsize\@plus-\hsize\@minus.01\hsize 
       \hskip 0em\@plus .3em
      \endgroup\fi
      \tnote{##1}\,\ignorespaces}%
   \let\TPToverlap\relax
   \def\endtablenotes{\par}%
}

\begin{document}

\renewcommand{\theenumi}{\alph{enumi}}
\setlength{\abovedisplayskip}{4pt plus 1pt minus 1pt}
\setlength{\belowdisplayskip}{4pt plus 0pt minus 1pt}
\setlength{\abovecaptionskip}{0pt plus 1pt minus 1pt}
\setlength{\belowcaptionskip}{0pt plus 0pt minus 1pt}
\setlength{\textfloatsep}{2pt plus 1pt minus 0pt}
\setlength{\floatsep}{3pt plus 0pt minus 1pt}
\setlength{\dbltextfloatsep}{3pt plus 1pt minus 1pt}
\setlength{\dblfloatsep}{3pt plus 0pt minus 1pt}

%
\title{A 1.1-$\:$/$\:$0.9-nA Temperature-Independent\\
213-$\:$/$\:$565-ppm/$^\circ$C Self-Biased CMOS-Only Current\\
Reference in 65-nm Bulk and 22-nm FDSOI}
%
%
%

\author{Martin~Lefebvre,~\IEEEmembership{Graduate Student Member,~IEEE},
		Denis~Flandre,~\IEEEmembership{Senior Member,~IEEE},
        and David~Bol,~\IEEEmembership{Senior Member,~IEEE}
\vspace{-0.75cm}
\thanks{This work was supported by the Fonds de la Recherche Scientifique (FRS-FNRS) of Belgium under grant CDR J.0014.20. \textit{(Corresponding author: Martin Lefebvre.)}
The authors are with the ICTEAM Institute, Université catholique de Louvain (UCLouvain), 1348 Louvain-la-Neuve, Belgium (e-mail: \{martin.lefebvre; denis.flandre; david.bol\}@uclouvain.be).
Color versions of one or more figures in this article are available at https://doi.org/10.1109/JSSC.2023.3240209.
Digital Object Identifier 10.1109/JSSC.2023.3240209
}}

%
%

\markboth{IEEE Journal of Solid-State Circuits,~Vol.~xx, No.~xx, xx~2022}%
{Shell \MakeLowercase{\textit{et al.}}: Bare Demo of IEEEtran.cls for IEEE Journals}
%
\IEEEoverridecommandlockouts
\IEEEpubid{\begin{minipage}{\textwidth}\ \\[12pt] \begin{scriptsize}This document is the paper as accepted for publication in JSSC, the fully edited paper is available at https://ieeexplore.ieee.org/document/10040689. \copyright 2023 IEEE. Personal use of this material is permitted. Permission from IEEE must be obtained for all other uses, in any current or future media, including reprinting/republishing this material for advertising or promotional purposes, creating new collective works, for resale or redistribution to servers or lists, or reuse of any copyrighted component of this work in other works.\end{scriptsize}
\end{minipage}} 
\maketitle

\begin{abstract} In many applications, the ability of current references to cope with process, voltage and temperature (PVT) variations is critical to maintain system-level performance. However, temperature-independent current references operating in the nA range are rarely area-efficient due to the use of large resistors which occupy a significant silicon area at this current level. In this paper, we introduce a nA-range constant-with-temperature (CWT) current reference relying on a self-cascode \mbox{MOSFET} (SCM), biased by a proportional-to-absolute-temperature voltage with a CWT offset. On the one hand, the proposed reference has been simulated post-layout in \mbox{65-nm} bulk. This design consumes 5.4~nW at 0.7~V and achieves a \mbox{1.1-nA} current with a line sensitivity (LS) of 0.69~$\%$/V and a temperature coefficient (TC) of 213~ppm/$^\circ$C. On the other hand, the proposed reference has been simulated and fabricated in \mbox{22-nm} fully-depleted silicon-on-insulator (FDSOI). This second design requires additional features to mitigate the impact of parasitic diode leakage at high temperature. In measurement, it consumes 5.8~nW at 0.9~V and achieves a \mbox{0.9-nA} current with a \mbox{0.39-$\%$/V} LS and a \mbox{565-ppm/$^\circ$C} TC. As a result of using an SCM, the proposed references occupy a silicon area of 0.0021~mm$^2$ in 65~nm (resp. 0.0132~mm$^2$ in 22~nm) at least 25$\times$ (resp. 4$\times$) smaller than state-of-the-art CWT references operating in the same current range.
\end{abstract}


%
\IEEEpeerreviewmaketitle

\vspace{-0.25cm}
\section{Introduction}
\label{sec:1_introduction}
\begin{figure}[!t]
	\centering
	\includegraphics[width=.49\textwidth]{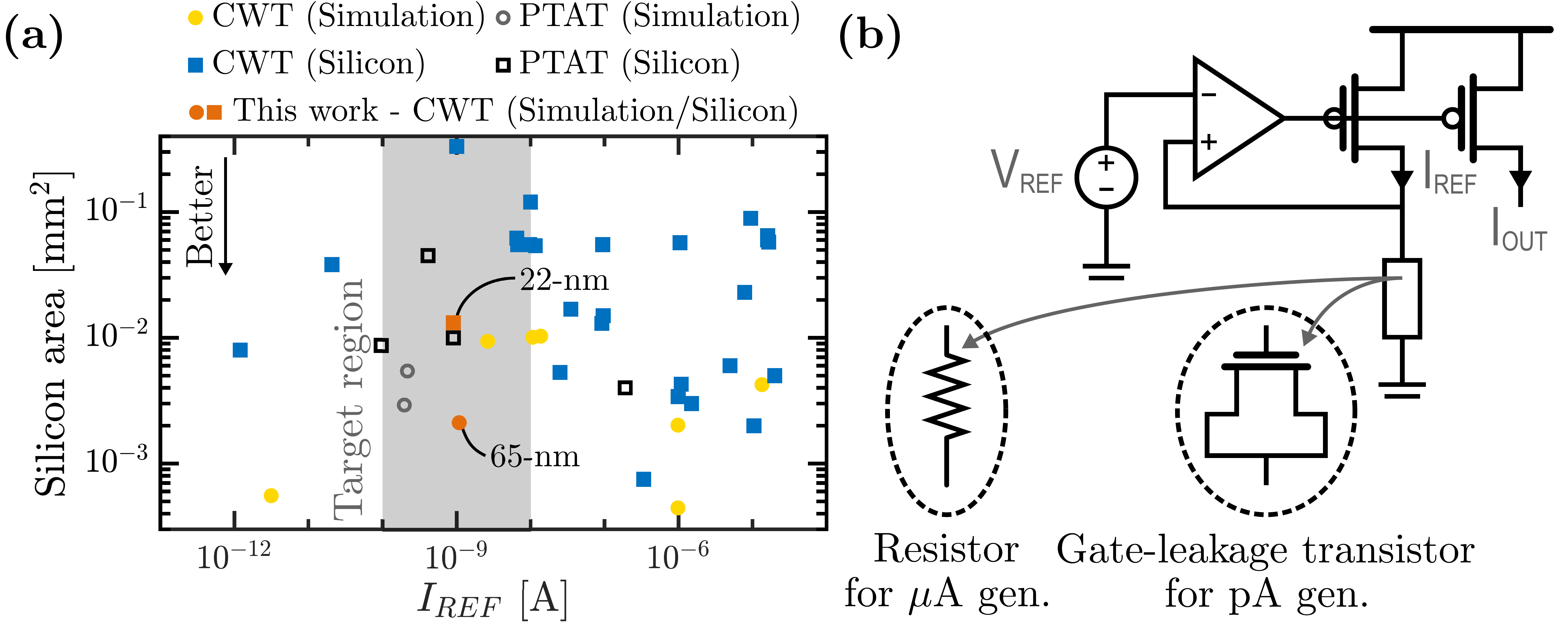}
	\caption{(a) Trade-off between silicon area and reference current, based on prior art, highlighting the absence of area-efficient solutions for the generation of a nA-range CWT current. (b) Conventional CWT current references are based on a reference voltage applied to a gate-leakage transistor or a resistor, which are respectively well suited to the generation of a pA- or $\mu$A-range current.}
	\label{fig:1_context}
\end{figure}

%
%
%
%
\IEEEPARstart{D}{esigning} current references to bias analog blocks constituting Internet-of-Things (IoT) nodes, such as low-power operational amplifiers \cite{Magnelli_2013} and wake-up timers \cite{Jeong_2015}, requires these references to be both area-efficient and robust to process, voltage and temperature (PVT) variations. This task appears to be of particular difficulty in the nA range, with Fig.~\ref{fig:1_context}(a) highlighting the absence of area-efficient solutions for the generation of a constant-with-temperature (CWT) nA-range current. This gap originates from the fact that conventional CWT current references [Fig.~\ref{fig:1_context}(b)] consist in applying a reference voltage to a voltage-to-current converter, which is usually implemented by a gate-leakage transistor \cite{Wang_2016, Wang_2018, Zhuang_2020} or a resistor \cite{Wang_2019_TCAS, Huang_2020}. These converters are well suited to the generation of a pA- or $\mu$A-range current, in the sense that they occupy a reasonable silicon area at this current level. Unfortunately, the generation of a nA-range current would lead to a much larger area, as the resistance of the gate-leakage transistor (resp. resistor) needs to be decreased (resp. increased) by placing devices in parallel (resp. series). Besides, proportional-to-absolute-temperature (PTAT) current references operating in the nA range, such as \cite{CamachoGaleano_2005, CamachoGaleano_2008}, are area-efficient because they rely on a self-cascode MOSFET (SCM) biased with a PTAT voltage. This structure generates a current proportional to the so-called specific sheet current and thus, to temperature. Even though prior designs have tried to harness the area efficiency of SCMs to generate a CWT current \cite{Huang_2010, Wang_2019_VLSI}, they cannot easily be ported to other technologies as they rely on a particular temperature dependence of the specific sheet current. An even more important challenge is thus to tackle key criteria such as the temperature coefficient (TC) and line sensitivity (LS) in the specific case of current references operating in the nA range.\looseness=-1\\
\IEEEpubidadjcol
\indent In this work, we propose a nA-range CWT current reference based on an SCM, which differs from \cite{CamachoGaleano_2005, CamachoGaleano_2008} by the addition of a CWT offset to the PTAT voltage biasing the SCM. This offset voltage is key in making SCM-based references CWT, and is obtained as the threshold voltage difference $\Delta V_T$ between two transistors of the same $V_T$ type, one of them being forward body-biased to reduce its threshold voltage. For the sake of generalization, this work expands on our conference paper \cite{Lefebvre_2022_ESSCIRC} by demonstrating the feasibility of the proposed reference in a conventional bulk technology, compared to fully-depleted silicon-on-insulator (FDSOI) only in \cite{Lefebvre_2022_ESSCIRC}. The proposed reference is validated in two CMOS technologies through post-layout simulations in TSMC \mbox{65-nm} bulk and measurements in GlobalFoundries (GF) \mbox{22-nm} FDSOI. In addition, it provides additional insight into the design, architecture and post-layout simulations of the reference in both technologies. The paper is structured as follows. Section~\ref{sec:2_governing_equations_and_operation_principle} explains the operation principle of the proposed reference based on its governing equations. Next, Section~\ref{sec:3_design_and_sizing_methodology} details the methodology used to size the reference. Then, Section~\ref{sec:4_optimization_and_simulation_results} discusses design optimizations, as well as simulation results in 65 and 22~nm, while Section~\ref{sec:5_measurement_results} presents the measurement results in 22~nm. Finally, Section~\ref{sec:6_comparison_to_the_state_of_the_art} compares the proposed reference to the state of the art, and Section~\ref{sec:7_conclusion} offers some concluding remarks.\looseness=-1

\section{Governing Equations and Operation Principle}
\label{sec:2_governing_equations_and_operation_principle}
\indent In this section, our first objective is to explain how the addition of a CWT offset to the voltage biasing the SCM leads to a temperature-independent reference current, based on general equations governing the current reference behavior. Then, we highlight the differences with respect to PTAT references \cite{CamachoGaleano_2005, CamachoGaleano_2008} and detail how temperature independence is achieved in the proposed reference with generic parameters. Finally, we compare different ways of implementing the offset voltage.\looseness=-1
\begin{figure}[!t]
	\centering
	\includegraphics[width=.488\textwidth]{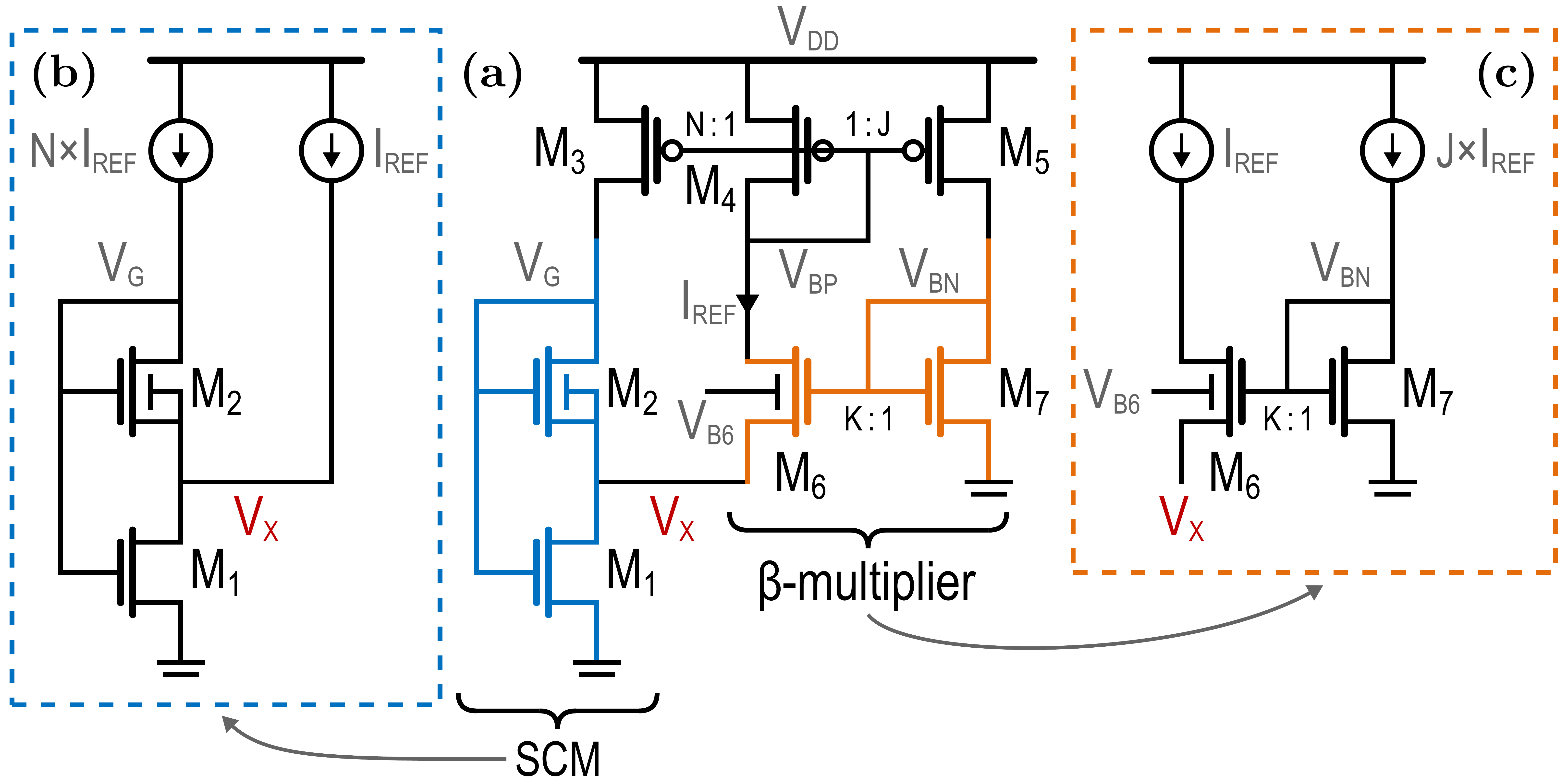}
	\caption{Basic schematic of the proposed current reference, which consists of an SCM formed by $M_{1-2}$ (in blue) and further simplified in (b), and a $\beta$-multiplier composed of $M_{6-7}$ (in orange) summarized in (c).}
	\label{fig:2_basic_schematic}
\end{figure}

\vspace{-0.25cm}
\subsection{Governing Equations}
\label{subsec:2A_governing_equations}
\indent To establish the equations governing the reference, we rely on the schematic in Fig.~\ref{fig:2_basic_schematic}(a), which consists of two main elements: an SCM and a $\beta$-multiplier, both supplied by pMOS current mirrors. First, the SCM relies on long-channel transistors in moderate inversion, for which a simplified model such as the analog compact MOSFET (ACM) model \cite{Cunha_1998} is adequate to describe the transistor current-voltage (\mbox{I-V}) curve. In this model, the drain current is given by
\begin{equation}
	I_D = I_{SQ}S(i_f-i_r)\textrm{,}\label{eq:id_acm}
\end{equation}
where $I_{SQ} = \frac{1}{2}\mu C_{ox}^{'} nU_T^2$ is the specific sheet current, $\mu$ is the carrier mobility, $C_{ox}^{'}$ is the normalized gate oxide capacitance, $n$ is the subthreshold slope factor, $U_T$ is the thermal voltage, $S = W/L$ is the transistor aspect ratio, and $i_f$, $i_r$ are the forward and reverse inversion levels. The transistor I-V curve is captured by
\begin{equation}
	V_P - V_S = U_T \left[\sqrt{1+i_f} - 2 + \log\left(\sqrt{1+i_f}-1\right)\right]\textrm{,}\label{eq:vs_acm}
\end{equation}
where $V_P = (V_G - V_{T0})/n$ is the pinch-off voltage, $V_{T0}$ is the threshold voltage at zero $V_{BS}$, and all voltages are referred to the transistor's body. A similar equation is obtained for the reverse inversion level, by replacing $i_{f}$ by $i_{r}$ and $V_S$ by $V_D$ in (\ref{eq:vs_acm}), but it is only relevant when the transistor is not saturated.\\
\indent Two distinct expressions of voltage $V_X$ are obtained by applying the ACM equations to transistors $M_{1-2}$ forming the SCM, and are given by (\ref{eq:vx_SCM_if1}) and (\ref{eq:vx_SCM_ir1}) in Appendix \ref{sec:8_appendix} with $\alpha \triangleq i_{f1}/i_{f2} > 1$ and $\beta \triangleq i_{r1}/i_{f2} \in [0;1]$. Then, the ratio of the aspect ratios of $M_{1-2}$ must comply with
\begin{equation}
	\frac{S_1}{S_2} = \frac{I_{SQ2}}{I_{SQ1}} \frac{1+N}{N} \frac{1}{\alpha - \beta}\label{eq:S1_over_S2}
\end{equation}
to ensure that Kirchhoff's current law is respected, $N$ being the current ratio between $M_3$ and $M_4$. Next, the ACM equations are applied to the weak-inversion transistors $M_{6-7}$ constituting the $\beta$-multiplier [Fig.~\ref{fig:2_basic_schematic}(c)]. We consider two distinct cases to obtain the expression of voltage $V_X$. First, the case in which $V_{B6} = V_X$ and $M_{6-7}$ have different threshold voltages. In this case, $V_X$ is given by
\begin{IEEEeqnarray}{RCL}
	V_X & \simeq & n_6 U_T \log(JK) + \left(\frac{n_6}{n_7}V_{T07} - V_{T06}\right)\textrm{,}\label{eq:vx_beta_mult_vt_type}\\
	& = & n U_T \log(JK) + \left(V_{T07} - V_{T06}\right)\textrm{,}\label{eq:vx_beta_mult_length}
\end{IEEEeqnarray}
$J$ and $K$ being current mirror ratios, with (\ref{eq:vx_beta_mult_vt_type}) corresponding to transistors with different $V_T$ types and hence entirely distinct characteristics, and (\ref{eq:vx_beta_mult_length}) to transistors of the same $V_T$ type but with different lengths, and thus similar characteristics except for their $V_{T0}$.
Second, we deal with the case in which $V_{B6} > V_X$, $M_{6-7}$ have the same $V_T$ type and length, and forward body$\:$/$\:$back-gate biasing (FBB) is leveraged to reduce $V_{T6}$. If the impact of the body effect on threshold voltage is modeled as $V_T = V_{T0} - \gamma_b V_{BS}$ for an nMOS, with $\gamma_b$ denoting the body factor, voltage $V_X$ is computed as
\begin{equation}
	V_X = n U_T \log(JK)+ \gamma_b\left(V_{B6} - V_X\right)\textrm{.}\label{eq:vx_beta_mult_fbb}
\end{equation}
Ultimately, regardless of the origin of $\Delta V_T$, (\ref{eq:vx_beta_mult_vt_type}) to (\ref{eq:vx_beta_mult_fbb}) boil down to a single expression
\begin{equation}
	V_X = n U_T \log(K_{PTAT}) + \Delta V_T\textrm{,}\label{eq:vx_beta_mult_final}
\end{equation}
with $K_{PTAT} \triangleq JK$. Connecting the SCM and the $\beta$-multiplier as in Fig.~\ref{fig:2_basic_schematic}(a) amounts to equating (\ref{eq:vx_SCM_if1}) and (\ref{eq:vx_beta_mult_final}), and leads to
\begin{IEEEeqnarray}{C}
	\left[\left(\sqrt{1+\alpha i_{f2}}-\sqrt{1+i_{f2}}\right) + \log\left(\frac{\sqrt{1+\alpha i_{f2}}-1}{\sqrt{1+i_{f2}}-1}\right)\right]\IEEEnonumber\\
	= \log(K_{PTAT}) + \textcolor{ECS-Red}{\frac{\Delta V_T}{n U_T}}\textrm{,}\label{eq:vx_equality}
\end{IEEEeqnarray}
in which only the term highlighted in red is temperature-dependent due to the combined temperature dependence of $\Delta V_T$ and $U_T$. Besides, applying (\ref{eq:id_acm}) to $M_2$ gives the expression of the reference current
\begin{equation}
	I_{REF}(T) = I_{SQ2}(T) i_{f2}(T) (S_2/N)\textrm{,}\label{eq:iref}
\end{equation}
where $I_{SQ2}(T) \propto U_T^2\mu(T) \propto T^{2-m}$, with $\mu(T) = \mu(T_0)\left(T/T_0\right)^{-m}$, and $m$ is the temperature exponent of the carrier mobility. An important quantity which remains to be defined is the sensitivity of the reference current to $V_X$,
\begin{equation}
	S_{I_{REF}} = \frac{1}{I_{REF}} \frac{dI_{REF}}{dV_X} = \frac{1}{I_{REF}} \frac{dI_{REF}}{di_{f2}} \frac{di_{f2}}{dV_X}\textrm{,}
\end{equation}
with $I_{REF}$ expressed by (\ref{eq:iref}) and $di_{f2}/dV_X$ computed from (\ref{eq:vx_SCM_if1}), consequently yielding
\begin{equation}
	S_{I_{REF}} = \frac{2}{i_{f2}n_2 U_T}\left[\frac{\alpha}{\sqrt{1+\alpha i_{f2}}-1} - \frac{1}{\sqrt{1+i_{f2}}-1}\right]^{-1}\textrm{.}\label{eq:siref}
\end{equation}
\indent In what follows, the LS and TC are computed using the box method, i.e.,
\begin{IEEEeqnarray}{RCL}
	\textrm{LS} & = & \frac{\left(I_{REF,\mathrm{max}}-I_{REF,\mathrm{min}}\right)}{I_{REF,\mathrm{avg}} \left(V_{DD,\mathrm{max}}-V_{DD,\mathrm{min}}\right)} \times 100 \: \%\textrm{/V,}\label{eq:LS}\\
	\textrm{TC} & = & \frac{\left(I_{REF,\mathrm{max}}-I_{REF,\mathrm{min}}\right)}{I_{REF,\mathrm{avg}} \left(T_{\mathrm{max}}-T_{\mathrm{min}}\right)} \times 10^6 \: \textrm{ppm/}^\circ\textrm{C,}\label{eq:TC}%
\end{IEEEeqnarray}
where $I_{REF,\mathrm{min/avg/max}}$ respectively stand for the minimum, average, and maximum reference current among the considered range. $V_{DD,\mathrm{min/max}}$ (resp. $T_{\mathrm{min/max}}$) refer to the lower and upper bounds of the voltage (resp. temperature) range.

\vspace{-0.25cm}
\subsection{Operation Principle}
\label{subsec:2B_operation_principle}
\begin{figure}[!t]
	\centering
	\includegraphics[width=.48\textwidth]{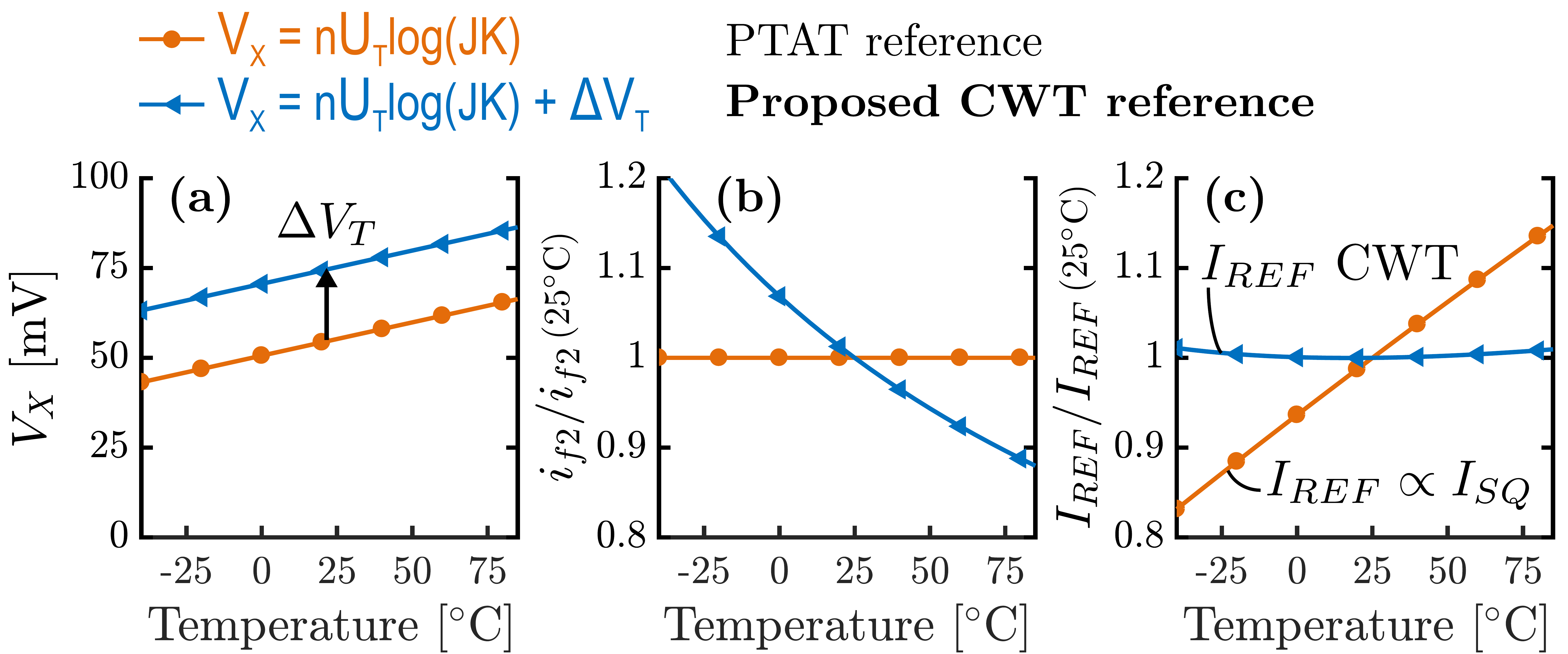}
	\caption{Operation principle of PTAT references proposed in prior art \cite{CamachoGaleano_2005, CamachoGaleano_2008} (in orange) and of the proposed CWT reference (in blue). Analytical expression of (a) the voltage $V_X$ applied to the SCM, (b) the inversion level of $M_2$, denoted as $i_{f2}$, and (c) the reference current $I_{REF}$, as a function of temperature and for $\Delta V_T$~=~20~mV. Generic technological parameters $n$~=~1.2 and $m$~=~1.25 are selected. (b) and (c) are normalized by their value at 25$^\circ$C. For the proposed CWT reference, the parameters leading to a minimum $I_{REF}$ TC are $K_{PTAT} = 6$ and $\alpha = 2.9$.}
	\label{fig:3_operation_principle}
\end{figure}
\begin{figure}[!t]
	\centering
	\includegraphics[width=.45\textwidth]{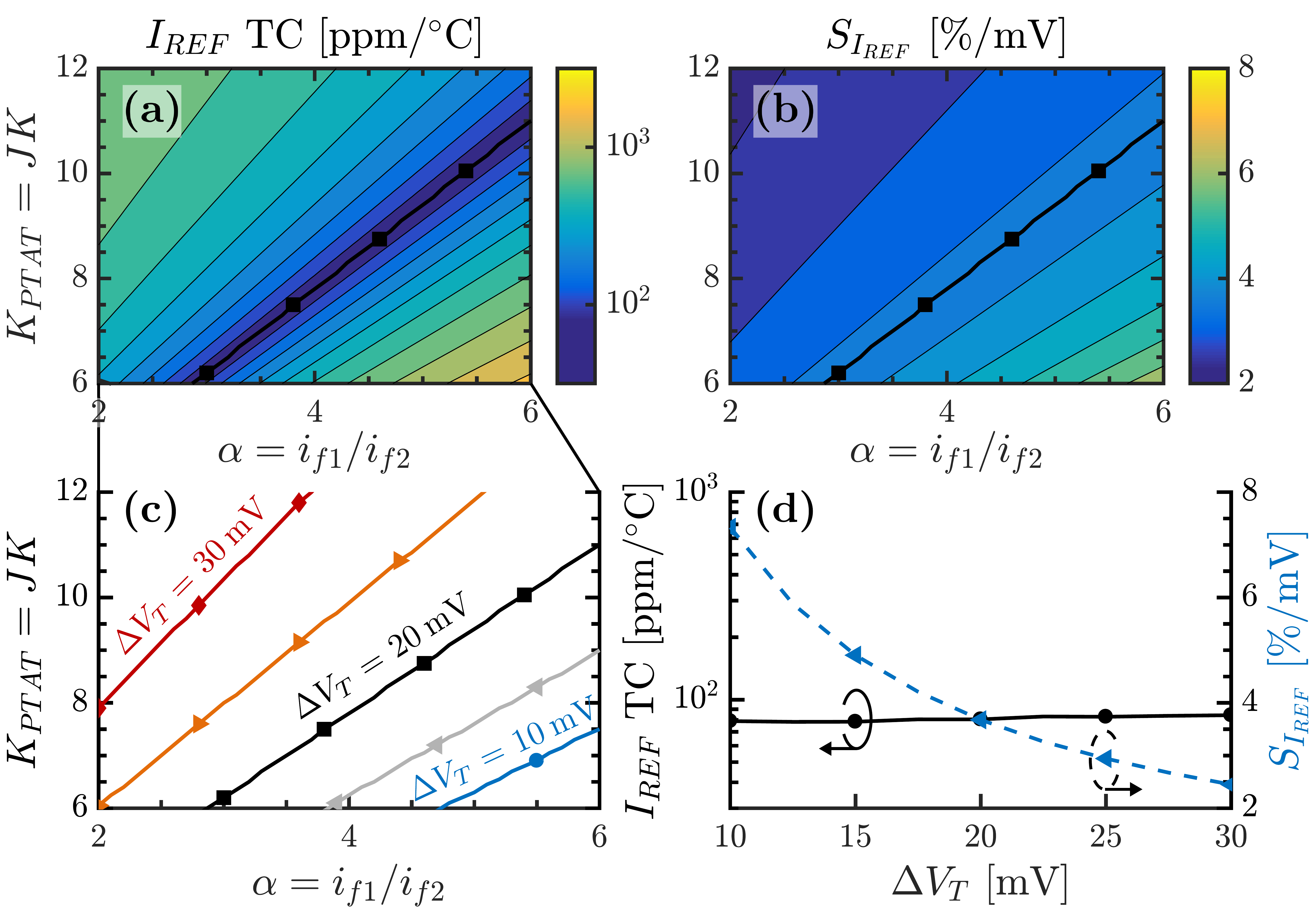}
	\caption{Analytical expression for (a) the TC of $I_{REF}$ and (b) the sensitivity $S_{I_{REF}}$, as a function of $K_{PTAT}$ and $\alpha$, and for the same technological parameters as Fig.~\ref{fig:3_operation_principle}. (c) Location of the $I_{REF}$ TC valley in the $(K_{PTAT};\:\alpha)$ space, for $\Delta V_T$ ranging from 10 to 30~mV. (d) $I_{REF}$ TC and $S_{I_{REF}}$ as a function of $\Delta V_T$, for $\alpha = 4$ and the optimal value of $K_{PTAT}$.}
	\label{fig:4_parameters_impact}
\end{figure}
In prior art \cite{CamachoGaleano_2005, CamachoGaleano_2008}, a PTAT reference current is obtained with $ V_{B6} = V_X$ and $M_{6-7}$ sharing the same $V_T$ type and length, thus yielding $\Delta V_T = 0$ and resulting in a purely PTAT voltage $V_X$ [Fig.~\ref{fig:3_operation_principle}(a)]. Equation~(\ref{eq:vx_equality}) subsequently becomes temperature-independent as the term in red amounts to zero. It also means that $i_{f2}$ does not change with temperature [Fig.~\ref{fig:3_operation_principle}(b)]. Ultimately, this forces $I_{REF}$ to have the same temperature dependence as $I_{SQ}$, as stated by (\ref{eq:iref}), and to follow a PTAT trend dictated by $T^{2-m}$ [Fig.~\ref{fig:3_operation_principle}(c)].\\
\indent In the proposed CWT current reference, an offset voltage $\Delta V_T > 0$ which is made CWT by design is added to $V_X$ [Fig.~\ref{fig:3_operation_principle}(a)], augmenting (\ref{eq:vx_equality}) with the temperature-dependent term in red. This causes $i_{f2}$ to decrease with temperature [Fig.~\ref{fig:3_operation_principle}(b)] and, with a proper selection of parameters $(K_{PTAT};\:\alpha)$, can make the temperature dependence of $i_{f2}$ compensate that of $I_{SQ}$. This results in a temperature-independent $I_{REF}$ as depicted in Fig.~\ref{fig:3_operation_principle}(c). An obvious question which arises at this stage is how to choose $(K_{PTAT};\:\alpha)$ to reach temperature independence. In Figs.~\ref{fig:4_parameters_impact}(a) and (b), we study their impact on two figures of merit (FoMs) of the current reference, namely the TC of $I_{REF}$ and $S_{I_{REF}}$, for a fixed CWT $\Delta V_T$~=~20~mV, and generic technological parameters $n$~=~1.2 and $m$~=~1.25. Fig.~\ref{fig:4_parameters_impact}(a) reveals the existence of an $I_{REF}$ TC valley for a linear relationship between $K_{PTAT}$ and $\alpha$. Furthermore, this valley corresponds to an iso-sensitivity curve in Fig.~\ref{fig:4_parameters_impact}(b). In Fig.~\ref{fig:4_parameters_impact}(c), we assess the influence of $\Delta V_T$ on the location of the $I_{REF}$ TC valley, and notice that increasing $\Delta V_T$ raises both the slope and offset of the linear relation between these two parameters. Technological parameters also have an influence on the valley location, with $n$ and $m$ having respectively an effect opposite and similar to $\Delta V_T$. However, these parameters are fixed by the technology and we do not consider them as potential tuning knobs for design. Lastly, Fig.~\ref{fig:4_parameters_impact}(d) depicts the evolution of the FoMs for $\Delta V_T$ ranging from 10 to 30~mV, $\alpha = 4$, and $K_{PTAT}$ corresponding to the $I_{REF}$ TC valley. It emphasizes that a constant \mbox{80-ppm/$^\circ$C} TC can be achieved with a proper choice of parameters, but that $S_{I_{REF}}$ improves from 7.33 to 2.45~$\%$/mV with a growing $\Delta V_T$. This trend is explained by the fact that an increase in $V_X$ pushes $M_{1-2}$ further into moderate inversion and decreases $S_{I_{REF}}$, at the cost of an increased area for the SCM and a larger minimum supply voltage. The values taken by parameters $(K_{PTAT};\:\alpha)$ are thus practically limited by such considerations during the sizing.

\vspace{-0.25cm}
\subsection{Practical Implementation of $\Delta V_{T}$}
\label{subsec:2C_practical_implementation_of_dvt}
The question that remains to be answered is how the offset voltage $\Delta V_T$ is practically achieved in the proposed current reference, knowing that it should ideally be CWT to ease the sizing, and that it does not need to be very large, as $S_{I_{REF}}$ improves less and less for large $\Delta V_T$ values [Fig.~\ref{fig:4_parameters_impact}(d)].
\begin{figure}[!t]
	\centering
	\includegraphics[width=.45\textwidth]{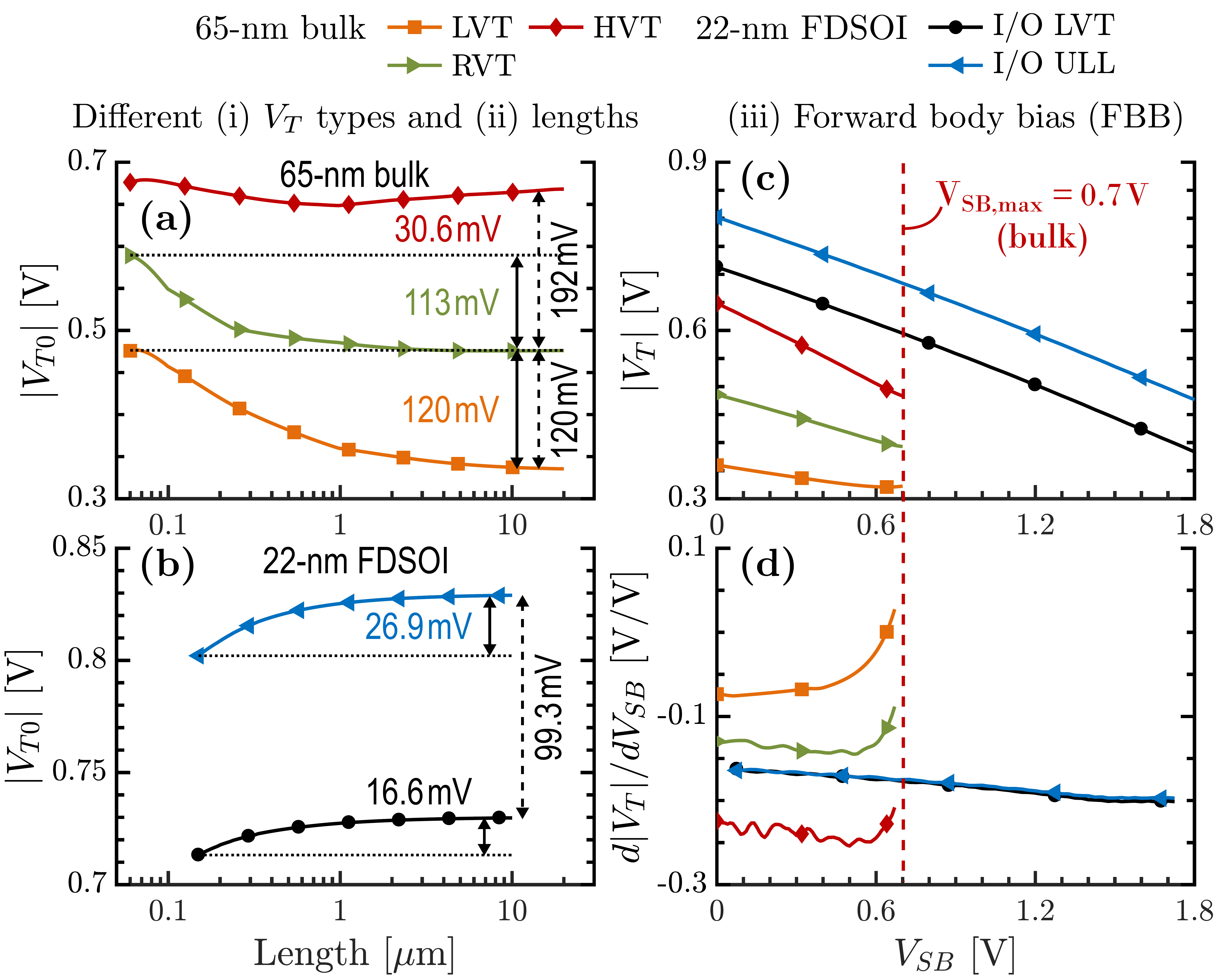}
	\caption{All figures correspond to \mbox{1-$\mu$m-wide} standard-well pMOS devices simulated at 25$^\circ$C, and the threshold voltage is extracted from $g_m/I_D$ vs. $V_{SG}$ curves as described in \cite{Jespers_2017}. $|V_{T0}|$ variations with the transistor length for (a) LVT, RVT and HVT core pMOS in \mbox{65-nm} bulk and (b) LVT and ULL I/O pMOS in \mbox{22-nm} FDSOI. For \mbox{1-$\mu$m-long} devices, (c) $|V_T|$ and (d) $d|V_T|/dV_{SB}$ with $V_{SB}$ for 65- and \mbox{22-nm} devices. $V_{SB}$ is limited by 0.7~V in 65~nm, to avoid the forward biasing of the parasitic diode between the transistor's source and body, and by the \mbox{1.8-V} supply voltage in 22~nm.}
	\label{fig:5_dvt_implementation}
\end{figure}
\begin{figure}[!t]
	\centering
	\includegraphics[width=.422\textwidth]{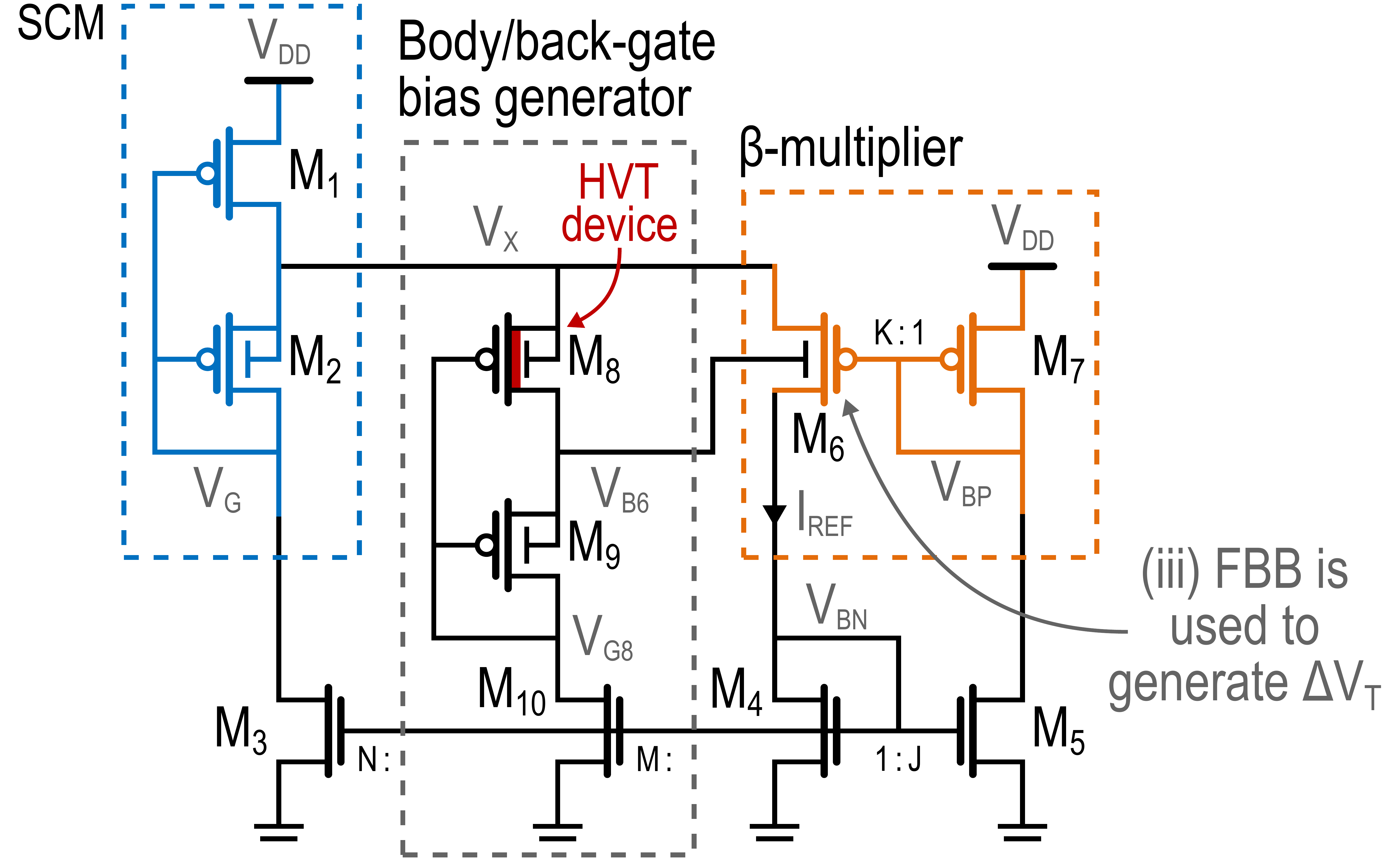}
	\caption{Schematic of the proposed reference with a pMOS-based topology, featuring a low-power voltage reference to generate the body$\:$/$\:$back-gate voltage $V_{B6}$. This voltage reference is formed by transistors $M_{8-10}$ and relies on the $V_T$ difference between HVT and LVT pMOS in \mbox{65-nm} bulk (resp. ULL and LVT pMOS in \mbox{22-nm} FDSOI). Note that an nMOS-based topology is also possible, at the cost of a slightly larger silicon area due to the use of triple-well devices.}
	\label{fig:6_pmos_implementation_schematic}
\end{figure}
In addition, a large $\Delta V_T$ also increases voltage $V_{GS1} = V_G$ in the SCM, resulting in a larger minimum supply voltage. A good trade-off is thus to have $\Delta V_T$ around 20 to 25~mV. Note that in what follows, we opt for an implementation based on a pMOS SCM as it limits the area overhead by using nwells instead of triple wells, and by leveraging the lower specific current of pMOS devices compared to nMOS ones. Three options are available to obtain a threshold voltage difference between two transistors: (i) different $V_T$ types, (ii) the same $V_T$ type but different lengths, and (iii) leveraging FBB to change the $V_T$ of one out of two transistors sharing the same $V_T$ type and length. Figs.~\ref{fig:5_dvt_implementation}(a) and (b) illustrate the variations of $|V_{T0}|$ for pMOS devices, respectively in TSMC \mbox{65-nm} bulk [Fig.~\ref{fig:5_dvt_implementation}(a)] in which \mbox{low-,} regular- and high-$V_T$ (LVT$\:$/$\:$RVT$\:$/$\:$HVT) core devices are available, and GF \mbox{22-nm} FDSOI [Fig.~\ref{fig:5_dvt_implementation}(b)], in which low-$V_T$ (LVT) and ultra-low-leakage (ULL) I/O devices are available. Both figures show that the $V_T$ difference resulting from using different transistor types is typically between 0.1 and 0.2~V, and is hence much larger than the $\Delta V_T$ we seek to achieve. Then, the same $V_T$ type but different lengths is a possibility to implement a $\Delta V_T$ in the desired range, with a difference from 31 to 120~mV in 65~nm, and from 17 to 27~mV in 22~nm. Nevertheless, this solution requires one of the transistors to be close to minimum length, which degrades other FoMs such as line sensitivity and mismatch, and leads to undesired second order effects due to the use of short-channel devices. Lastly, Figs.~\ref{fig:5_dvt_implementation}(c) and (d) represent the evolution of $V_T$ and the derivative $dV_T/dV_{SB}$ as a function of $V_{SB}$. Fig.~\ref{fig:5_dvt_implementation}(d) shows that in 65~nm, the body factor ranges from 74 to 244~mV/V depending on the $V_T$ type, while in 22~nm, it is around 165~mV/V close to zero $V_{SB}$, i.e., body$\:$/$\:$back-gate and source connected to the same voltage. In this work, we choose to exploit FBB as it allows to obtain a positive offset $\Delta V_T$ contrary to RBB, provides a sufficient tuning range for a \mbox{20-to-25-mV} $\Delta V_T$, and offers the advantage of having devices $M_{6-7}$ with the same technological parameters.\\
\indent A schematic of the proposed current reference including the body$\:$/$\:$back-gate bias generator ($M_{8-10}$) is represented in Fig.~\ref{fig:6_pmos_implementation_schematic}. Note that the topology has been inverted compared to Fig.~\ref{fig:2_basic_schematic}(a), i.e., nMOS and pMOS devices are swapped, as well as ground and supply voltage connections. The proposed reference is thus built around a pMOS-based SCM. The voltage reference is based on the $\Delta V_{SG}$ between an HVT ($M_8$) and LVT ($M_9$) devices biased by a current source ($M_{10}$). It generates a positive reference voltage thanks to the $V_{T}$ difference between $M_{8-9}$, which are practically implemented with core HVT and LVT devices in 65~nm (resp. I/O ULL and LVT devices in 22~nm). Assuming a weak-inversion transistor in saturation, i.e., with $V_{SD} > 4U_T$, $V_{SG}$ is expressed as
\begin{equation}
	V_{SG} = |V_{T0}| + nU_T\log\left(\frac{I_{SD}}{I_{SQ}S}\right)\textrm{,}\label{eq:vsg_subthreshold}
\end{equation}
where $I_{SQ} = \mu C_{ox}^{'}(n-1)U_T^2$ is another definition of the specific sheet current. Using (\ref{eq:vsg_subthreshold}), $V_{SB6}$ is equal to
\begin{equation}
	V_X-V_{B6} = \left|V_{T08}-V_{T09}\right| + nU_T\log\left(\frac{I_{SQ9}S_9}{I_{SQ8}S_8}\right)\textrm{,}\label{eq:vb6_minus_vx}
\end{equation}
in which the first term is either PTAT or complementary-to-absolute-temperature (CTAT) depending on the TC of the threshold voltages, and the second term's TC can be tuned by changing the ratio $S_9/S_8$. This voltage reference will be sized so that $\Delta V_T$ between $M_6$ and $M_7$ is temperature-independent, as detailed in Section~\ref{subsec:3B_body_back_gate_bias_generator}. Finally, the voltage reference changes the ratio (\ref{eq:S1_over_S2}) to
\begin{equation}
	\frac{S_1}{S_2} = \frac{I_{SQ2}}{I_{SQ1}} \frac{1+\mathbf{M}+N}{N} \frac{1}{\alpha - \beta}\textrm{.}\label{eq:S1_over_S2_final}
\end{equation}

\section{Design and Sizing Methodology}
\label{sec:3_design_and_sizing_methodology}
\subsection{Overview of the Methodology}
\begin{figure}[!t]
	\centering
	\includegraphics[width=.45\textwidth]{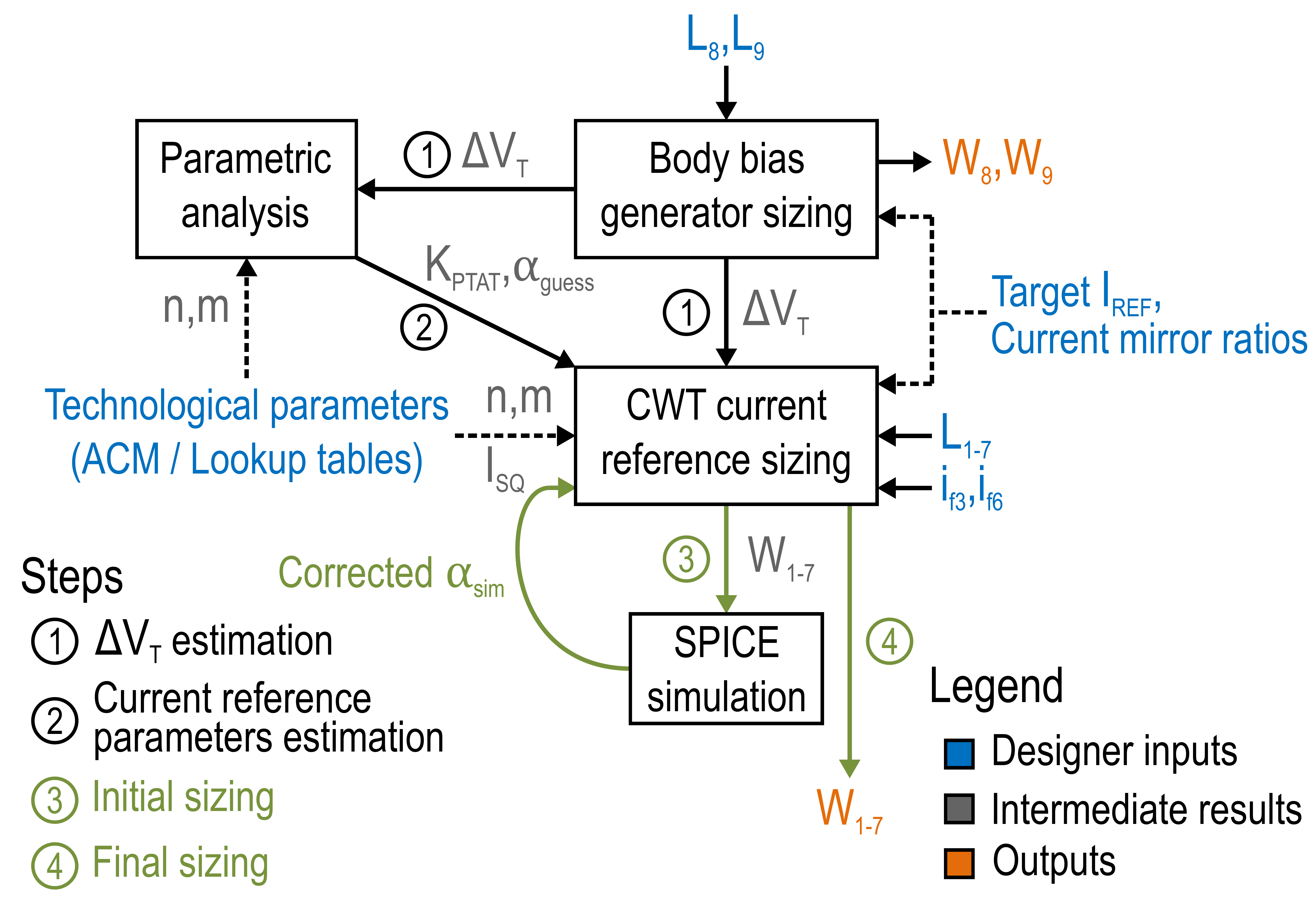}
	\caption{Four-step flowchart of the design and sizing methodology.}
	\label{fig:7_sizing_methodology}
\end{figure}
Fig.~\ref{fig:7_sizing_methodology} provides a flowchart of the sizing methodology. It requires inputs from the designer, i.e., a target reference current, current mirror ratios, transistor lengths and inversion levels, as well as technological parameters, and it outputs transistor widths. The methodology is divided into four steps.
\par\mbox{Step 1)} consists in sizing the body$\:$/$\:$back-gate bias generator by selecting $W_{8-9}$ such that $\Delta V_T$'s TC is minimized.
\par\mbox{Step 2)} makes an educated guess of the value of $\alpha$ minimizing the TC of $I_{REF}$ for a fixed $K_{PTAT}$, which is denoted as $\alpha_{\mathrm{guess}}$. The prediction is based on the linear relationship between $K_{PTAT}$ and $\alpha$ obtained through the parametric analysis developed in Section~\ref{subsec:2A_governing_equations}. It requires the $\Delta V_T$ computed by \mbox{step 1)}, and technological parameters $n$ and $m$, obtained by fitting the ACM model to $g_m/I_D$ curves extracted from SPICE simulations as lookup tables \cite{Jespers_2017}.
\par\mbox{Step 3)} sizes the current reference for a range of $\alpha$ values and a fixed $K_{PTAT}$, based on $\Delta V_T$ and parameters $n$, $m$ and $I_{SQ}$. Several sizings, which differ by the widths of $M_{1-2}$, are used to perform pre-layout SPICE simulations, from which we extract the value of $\alpha$ leading to a minimum $I_{REF}$ TC. This value is referred to as $\alpha_{\mathrm{sim}}$ and should be close to $\alpha_{\mathrm{guess}}$.
\par\mbox{Step 4)} simply consists in running the sizing algorithm with $\alpha = \alpha_{\mathrm{sim}}$, thereby delivering the final transistor widths.

\vspace{-0.25cm}
\subsection{Body$\:$/$\:$Back-Gate Bias Generator}
\label{subsec:3B_body_back_gate_bias_generator}
\begin{figure}[!t]
	\centering
	\includegraphics[width=.45\textwidth]{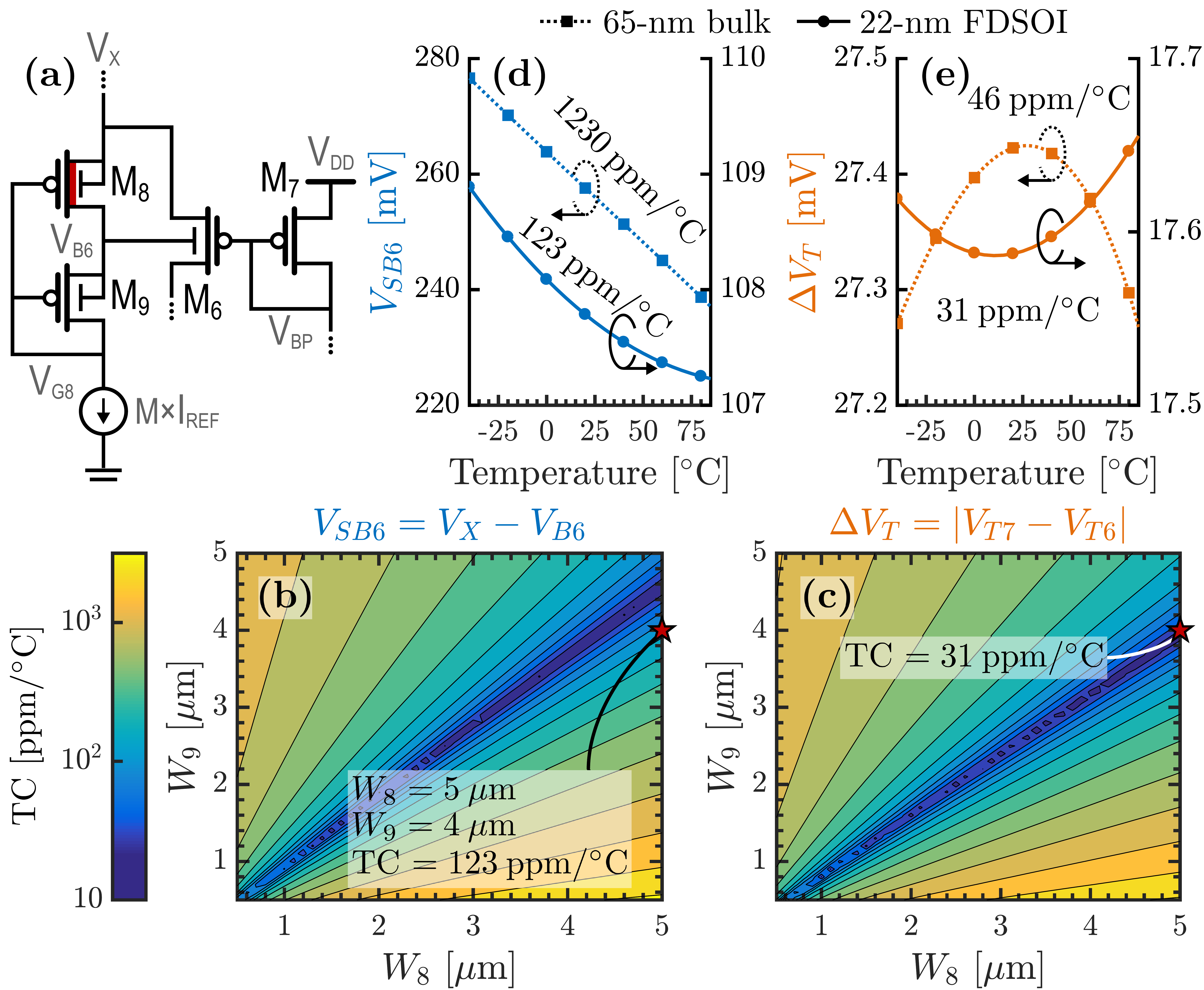}
	\caption{$\Delta V_T$ is made CWT by tuning the TC of $V_{SB6}$ by means of the ratio $S_9/S_8$ (\ref{eq:vb6_minus_vx}). All figures correspond to the behavior in TT. (a) Schematic and testbench of the low-power voltage reference used to generate $V_{B6}$. TC of (b) $V_X - V_{B6}$ and (c) $\Delta V_T$, for different widths of $M_{8-9}$ with $L$~=~8~$\mu$m in 22~nm. (d) Reference voltage $V_X - V_{B6}$ and (e) resulting threshold voltage difference $\Delta V_T$ as a function of temperature at the chosen design point.}
	\label{fig:8_sizing_body_bias_gen}
\end{figure}

The setup employed to size the body$\:$/$\:$back-gate bias generator is illustrated in Fig.~\ref{fig:8_sizing_body_bias_gen}(a). It is composed of a low-power voltage reference formed by $M_{8-9}$, biased by an ideal current source $M \times I_{REF}$ with $M$~=~1 and $I_{REF}$~=~1.25~nA, and transistors $M_{6-7}$, respectively body-biased by a positive and zero $V_{SB}$. For $M_{8-9}$, we choose a relatively long length of 5~$\mu$m in 65~nm and 8~$\mu$m in 22~nm to reduce LS and variability. In Figs.~\ref{fig:8_sizing_body_bias_gen}(b) and (c), the width of $M_{8-9}$ is swept from 0.5 to 5~$\mu$m in 0.1-$\mu$m steps to select the design point minimizing the TC of $\Delta V_T$ in 22~nm. Figs.~\ref{fig:8_sizing_body_bias_gen}(b) and (d) show that it corresponds to a slightly CTAT yet close to CWT $V_X - V_{B6}$, with a 123-ppm/$^\circ$C TC, and achieves a 31-ppm/$^\circ$C $\Delta V_T$ in Figs.~\ref{fig:8_sizing_body_bias_gen}(c) and (e). This result is consistent with the fact that $\Delta V_T$ is obtained through the back-gate effect, as described for an nMOS by $V_T(V_{BS}) = V_{T0} - \gamma_b V_{BS}$, where the body factor $\gamma_b$ is temperature-independent at first order in FDSOI~\cite{daSilva_2021}. In 65~nm, a CTAT 1230-ppm/$^\circ$C $V_X - V_{B6}$ voltage is required to obtain a CWT 46-ppm/$^\circ$C $\Delta V_T$, as it is obtained through the temperature-dependent body effect
\begin{equation}
	V_T (V_{BS}) = V_{T0} + \gamma_b \left(\sqrt{2\phi_{fp} - V_{BS}} - \sqrt{2\phi_{fp}}\right)\textrm{,}\label{eq:vt_body_effect}
\end{equation}
where $\phi_{fp}$ is Fermi's potential.

\vspace{-0.25cm}
\subsection{Temperature-Independent Current Reference}
\label{subsec:3C_temperature_independent_current_reference}
\begin{figure}[!t]
	\centering
	\includegraphics[width=.45\textwidth]{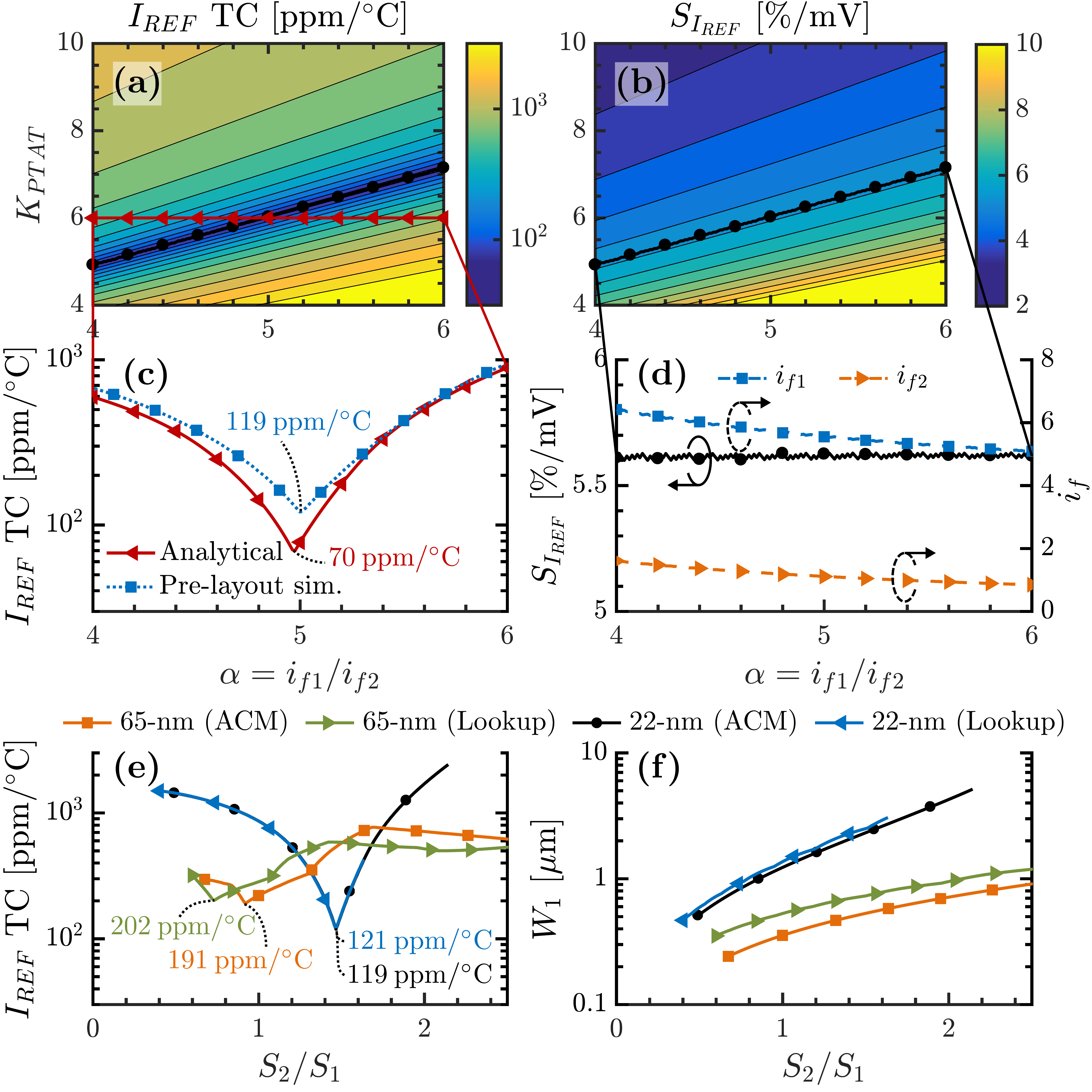}
	\caption{$I_{REF}$ is made CWT by properly selecting $K_{PTAT}$ and $\alpha$. (a) $I_{REF}$ TC and (b) $S_{I_{REF}}$ for different values of $(K_{PTAT};\:\alpha)$ with $\Delta V_T$~=~17.6~mV, $m$~=~1, and $n$~=~1.2, obtained from step~1) of the sizing methodology. In 22~nm and for different values of $\alpha$, (c) $I_{REF}$ TC computed from the analytical model and pre-layout simulation results, with the reference sized using an ACM-based approach, and (d) $S_{I_{REF}}$ and $i_{f1}$, $i_{f2}$. In 65 and 22~nm, (e) $I_{REF}$ TC and (f) $W_1$ as a function of $S_2/S_1$, for both ACM- and lookup-table-based sizings.}
	\label{fig:9_sizing_current_ref}
\end{figure}

The parametric analysis yields Figs.~\ref{fig:9_sizing_current_ref}(a) and (b), which are akin to Fig.~\ref{fig:4_parameters_impact}, but for 22~nm and a $\Delta V_T$ of 17.6~mV. As already described in Section~\ref{subsec:2B_operation_principle}, the $I_{REF}$ TC valley corresponds to a linear relationship between $K_{PTAT}$ and $\alpha$, along with an \mbox{iso-$S_{I_{REF}}$} curve. Next, the current reference sizing per se can be performed in two ways. The first one is similar to the sizing methodology proposed in \cite{CamachoGaleano_2005, CamachoGaleano_2008} and is based solely on the ACM model.
It consists of five main steps:
\begin{enumerate}
	\item Compute voltage $V_X$ using (\ref{eq:vx_beta_mult_final});
	\item Solve (\ref{eq:vx_SCM_if1}) for $i_{f2}$, then calculate $i_{f1} = \alpha i_{f2}$, $S_{I_{REF}}$ from (\ref{eq:siref}), and $S_2$ from (\ref{eq:id_acm});
	\item Solve (\ref{eq:vx_SCM_ir1}) for $\beta$, calculate $i_{r1} = \beta i_{f2}$ and $S_1$ from (\ref{eq:S1_over_S2_final});
	\item Compute the aspect ratio of $M_{6-7}$ using (\ref{eq:id_acm});
	\item Compute the aspect ratio of transistors $M_{3-5}$ and $M_{10}$ forming the current mirrors using (\ref{eq:id_acm}).
\end{enumerate}
The second way to perform the sizing replaces \mbox{steps b) -- c)} by a direct use of lookup tables. The gate voltage of $M_{1-2}$ is swept while keeping $V_X$ to the value fixed by a), thus providing sizes complying with the SCM structure. This second method is theoretically more accurate, as transistors are better described by lookup tables than by a limited set of parameters.\looseness=-1\\
\indent Let us now take a closer look at the results of the sizing algorithm, with Figs.~\ref{fig:9_sizing_current_ref}(a) to (d) presenting sizing results in 22~nm and Figs.~\ref{fig:9_sizing_current_ref}(e) and (f), results in both 65 and 22~nm. Fig.~\ref{fig:9_sizing_current_ref}(c) is obtained by fixing $K_{PTAT}$ to 6 and extracting a slice of the 3D surface in Fig.~\ref{fig:9_sizing_current_ref}(a). It reveals that the \mbox{70-ppm/$^\circ$C} TC found for $\alpha_{\mathrm{guess}}$~=~4.975 closely matches the location of the minimum TC obtained from pre-layout simulations for $\alpha_{\mathrm{sim}}$~=~5, but differs in terms of value with a slightly larger \mbox{119-ppm/$^\circ$C} TC. Moreover, Fig.~\ref{fig:9_sizing_current_ref}(d) indicates that, along the $I_{REF}$ TC valley, $S_{I_{REF}}$ remains stable around 5.62~$\%$/mV while $i_{f1}$, $i_{f2}$ decrease for larger values of $\alpha$, as $M_{1-2}$ are pushed closer to weak inversion. Then, Fig.~\ref{fig:9_sizing_current_ref}(e) depicts the TC of $I_{REF}$ computed from \mbox{pre-layout} simulations as a function of $S_2/S_1$, which is a proxy for $\alpha$ (\ref{eq:S1_over_S2_final}).  In~65~nm, pre-layout simulation results faintly diverge, with the ACM-based sizing leading to a \mbox{191-ppm/$^\circ$C} TC at $S_2/S_1$~=~0.92, and the lookup-table-based one to a \mbox{202-ppm/$^\circ$C} TC at $S_2/S_1$~=~0.73. In addition, the ACM-based sizing tends to output a narrower transistor $M_1$ than its lookup-table-based counterpart [Fig.~\ref{fig:9_sizing_current_ref}(f)]. These discrepancies are presumably related to the approximated transistor behavior resulting from the estimation of ACM model parameters, as well as second order effects such as channel width modulation. In~22~nm, both sizing methods come up with a minimum at $S_2/S_1$~=~1.47, and Fig.~\ref{fig:9_sizing_current_ref}(f) confirms that the sizes output by both algorithms are in close agreement.

\vspace{-0.25cm}
\section{Optimization and Simulation Results}
\label{sec:4_optimization_and_simulation_results}
This section comes in two parts. First, Sections~\ref{subsec:4A_line_sensitivity_improvement} and \ref{subsec:4B_parasitic_diode_leakage_mitigation} introduce architectural optimizations to the proposed reference, by respectively discussing how to improve LS and how to limit the TC degradation due to parasitic diode leakage at high temperature. Next, Section~\ref{subsec:4C_final_implementation_performance} presents the schematic and post-layout simulation results of the proposed reference implemented in two common scaled technologies, i.e., TSMC 65-nm bulk and GF 22-nm FDSOI.

\vspace{-0.25cm}
\subsection{Line Sensitivity Improvement by Low-Voltage Cascoding}
\label{subsec:4A_line_sensitivity_improvement}
\begin{figure}[!t]
	\centering
	\includegraphics[width=.478\textwidth]{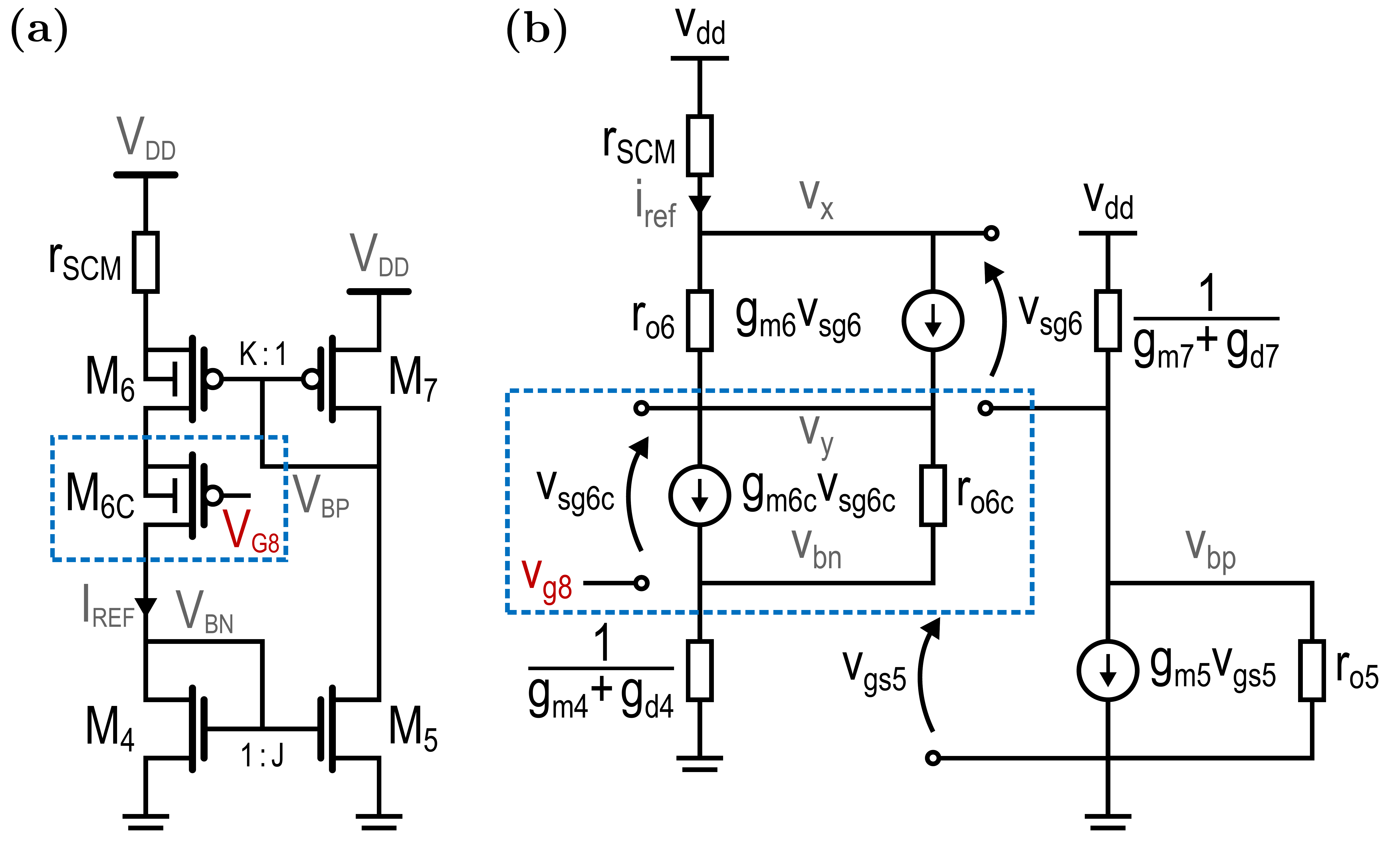}
	\caption{(a) Schematic of the $\beta$-multiplier with low-voltage cascoding of $M_6$ through transistor $M_{6C}$. (b) Small signal schematic of (a).}
	\label{fig:10_low_voltage_cascode}
\end{figure}

\begin{figure*}[!t]
	\centering
	\includegraphics[width=.9\textwidth]{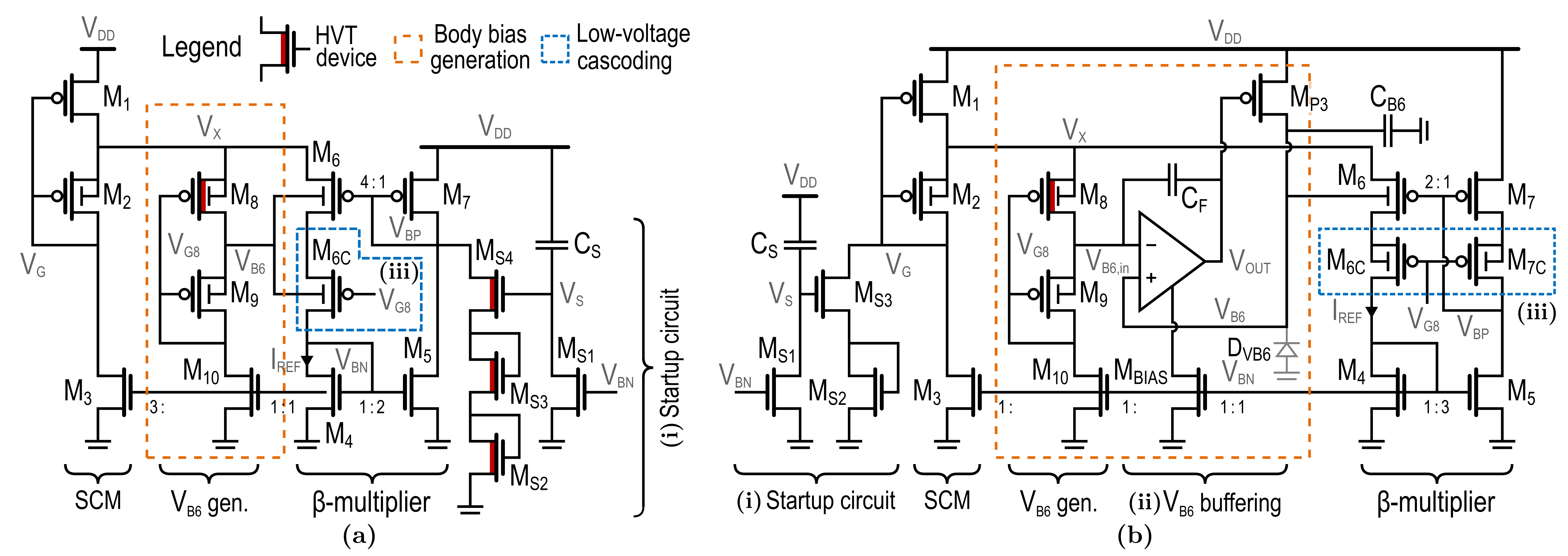}
	\caption{Complete schematic of the proposed nA-range CWT reference in (a) TSMC \mbox{65-nm} bulk and (b) GF \mbox{22-nm} FDSOI. The schematic includes a dynamic startup circuit similar to \cite{CamachoGaleano_2008}, a buffering stage of the body$\:$/$\:$back-gate voltage of $M_6$, and a low-voltage cascode. Note that the startup circuits are slightly different due to the presence of the buffering stage in 22~nm. In the \mbox{65-nm} design, $M_8$ and $M_{S2-S4}$ are implemented with core HVT pMOS devices, while in the \mbox{22-nm} one, $M_8$ is implemented with an I/O ULL pMOS devices.}
	\label{fig:14_final_implementation_schematic}
\end{figure*}

Low-voltage cascoding is often used in scaled technologies to improve LS without compromising the minimum supply voltage. An analytical approach allows to better grasp which transistors induce a dependence to supply voltage in the first place. The SCM in Fig.~\ref{fig:6_pmos_implementation_schematic} can be simplified by a resistor [Fig.~\ref{fig:10_low_voltage_cascode}(a)] with an equivalent behavior around the operation point, denoted as $Q$. For this pMOS implementation, the resistance is given by
\begin{equation}
	r_{SCM} = \left.\frac{\partial \left(V_{DD}-V_X\right)}{\partial I_{REF}}\right|_Q = \frac{1}{I_{REF} S_{I_{REF}}}\textrm{.}\label{eq:rSCM}
\end{equation}
\noindent Relying on the small signal schematic in Fig.~\ref{fig:10_low_voltage_cascode}(b) and first considering the case without transistor $M_{6C}$, i.e., $v_y = v_{bn}$, the absolute LS of $V_X$ and the relative LS of $I_{REF}$ around $Q$ can be expressed as
\begin{IEEEeqnarray}{RCL}
	\left.\frac{\partial \left(V_{DD}-V_X\right)}{\partial V_{DD}}\right|_Q & = &  \frac{v_{dd}-v_x}{v_{dd}} = \frac{\frac{g_{d5}}{J} + g_{d6}}{g_{m6}}\textrm{,}\label{eq:LS_small_sig_vx_basic}\\
	\frac{1}{I_{REF}} \left.\frac{\partial I_{REF}}{\partial V_{DD}}\right|_Q & = & \frac{1}{I_{REF}}\frac{i_{ref}}{v_{dd}} = S_{I_{REF}} \frac{v_{dd}-v_x}{v_{dd}}\textrm{,}\label{eq:LS_small_sig_iref}
\end{IEEEeqnarray}
under the common assumption that $g_m \gg g_d = 1/r_o$. Equation~(\ref{eq:LS_small_sig_vx_basic}) indicates that the dependence of $v_{dd}-v_x$ to $v_{dd}$ can be attributed to $M_{5-6}$'s output conductance. Next, $M_{6C}$ in Fig.~\ref{fig:10_low_voltage_cascode}(a) is used to cascode $M_6$, and is biased by voltage $V_{G8}$ originating from the body$\:$/$\:$back-gate bias generator [Fig.~\ref{fig:6_pmos_implementation_schematic}]. A complete low-voltage cascode as depicted in Fig.~\ref{fig:14_final_implementation_schematic}(b) can also be used, but does not bring any significant LS improvement as $M_7$ is in diode and already has negligible $V_{DS}$ variations. Based on Fig.~\ref{fig:10_low_voltage_cascode}(b) and assuming that $v_{g8} \simeq v_x$, the absolute LS of $V_X$ becomes
\begin{figure}[!t]
	\centering
	\includegraphics[width=.45\textwidth]{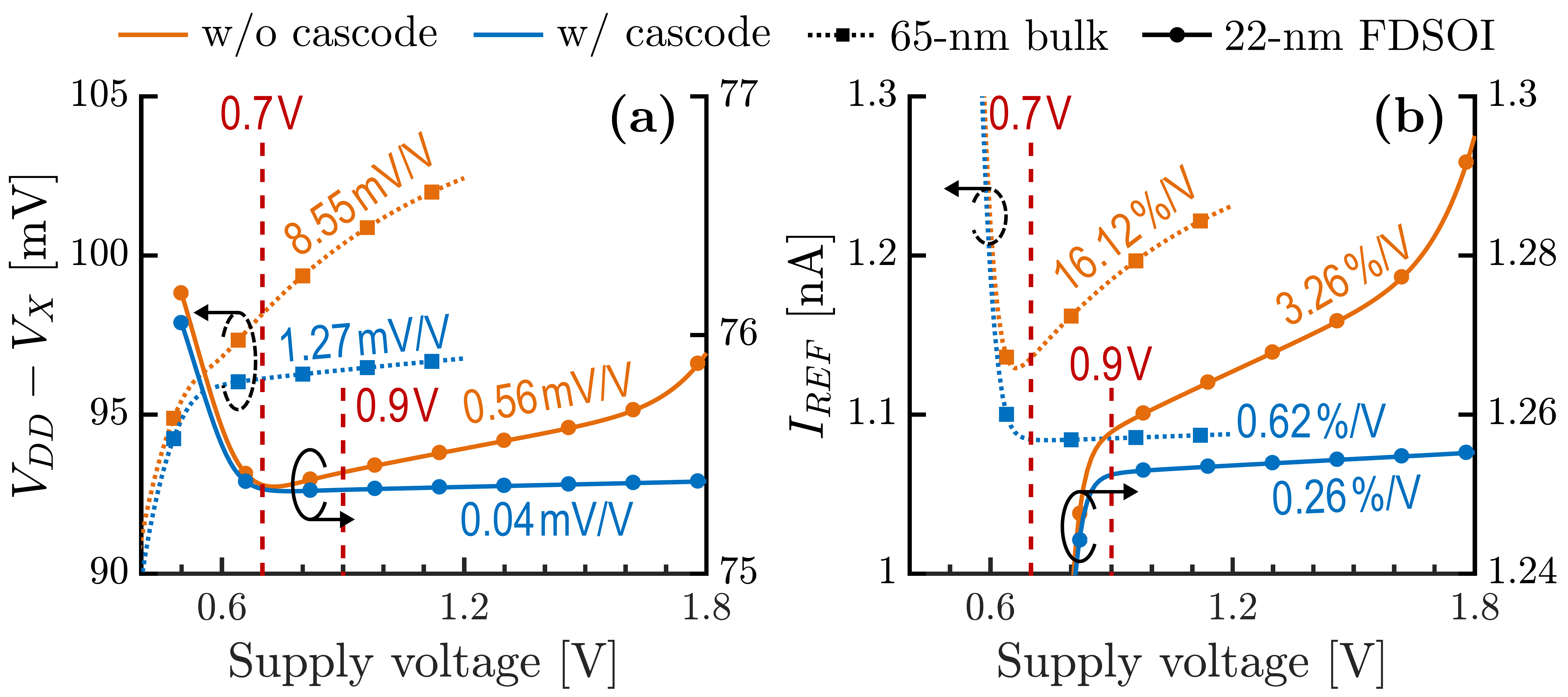}
	\caption{Pre-layout simulations of (a) $V_{DD}-V_X$ and (b) $I_{REF}$ with respect to the supply voltage in the TT 25$^\circ$C corner, highlighting the improvement of LS achieved by cascoding $M_6$ in both 65- and \mbox{22-nm} technologies. The minimum supply voltage is 0.7~V in 65~nm and 0.9~V in 22~nm.}
	\label{fig:11_low_voltage_cascode_sim}
\end{figure}
\begin{equation}
	\frac{v_{dd}-v_x}{v_{dd}} = \frac{\frac{g_{d5}}{J}}{g_{m6} + \left(\frac{g_{d6}}{g_{m6c}}\right)r_{SCM}} \simeq  \frac{\frac{g_{d5}}{J}}{g_{m6}}\textrm{,}\label{eq:LS_small_sig_vx_cascode}
\end{equation}
while the relative LS of $I_{REF}$ is still given by (\ref{eq:LS_small_sig_iref}). LS is reduced because $g_{d6}$ no longer appears in (\ref{eq:LS_small_sig_vx_cascode}), which is highly beneficial as $M_6$ is in weak inversion while $M_5$ is closer to moderate inversion, meaning that $M_6$ is shorter than $M_5$ and therefore has a larger output conductance. In addition, current mirrors are implemented using composite transistors, i.e., the series connection of devices sharing the same $V_G$ and $V_B$ \cite{Galup-Montoro_1994}, further reducing the output conductance.\\
\indent In 65~nm, simulations without cascoding exhibit an LS of $V_X$ and $I_{REF}$ [Fig.~\ref{fig:11_low_voltage_cascode_sim}] respectively worth 8.55~mV/V and 16.12~$\%$/V, computed from 0.7 to 1.2~V using the \textbf{box method}. Based on the small signal parameters extracted at the center of the supply voltage range, i.e., 0.95~V, (\ref{eq:LS_small_sig_vx_basic}) predicts an LS of $V_X$ of 7.81~mV/V which compares fairly well to the \mbox{8.25-mV/V} simulated value, computed with \textbf{first-order finite differences} around $Q$ rather than with the box method. Equation~(\ref{eq:LS_small_sig_iref}) links the LS of $I_{REF}$ to $v_{dd}-v_x/v_{dd}$, and yields a \mbox{21.33-$\%$/V} value diverging from the \mbox{15.75-$\%$/V} simulation, likely due to an overestimation of $S_{I_{REF}}$. With cascoding, the LS of $V_X$ and $I_{REF}$ improve to 1.27~mV/V and 0.62~$\%$/V. Besides, the \mbox{0.90-mV/V} LS of $V_X$ prediction agrees with the \mbox{1.25-mV/V} simulation. Similarly to the non-cascoded case, the \mbox{2.46-$\%$/V} prediction for the LS of $I_{REF}$ is exaggerated compared to the \mbox{0.83-$\%$/V} simulation.\\
\indent Simulations without cascoding in 22~nm yield a \mbox{0.56-mV/V} LS of $V_X$ and a \mbox{3.26-$\%$/V} LS of $I_{REF}$ from 0.9 to 1.8~V. The small signal approach estimates an LS of $V_X$ worth 0.35~mV/V around 1.35~V, which compares fairly well to the \mbox{0.33-mV/V} simulated value. A \mbox{1.96-$\%$/V} LS of $I_{REF}$ is predicted by (\ref{eq:LS_small_sig_iref}) and closely matches the \mbox{1.89-$\%$/V} simulation, thus confirming the correct evaluation of $S_{I_{REF}}$~=~5.85~$\%$/mV. Cascoding reduces LS to 0.04~mV/V and 0.26~$\%$/V for $V_X$ and $I_{REF}$. Moreover, predictions give a \mbox{0.04-mV/V} LS of $V_X$ which agrees with the 0.038-mV/V simulation, while the LS of $I_{REF}$ is accurately predicted to be 0.22~$\%$/V, which is close to the \mbox{0.23-$\%$/V} simulation as for the case without cascoding.\\
\indent To conclude, cascoding decisively diminishes $I_{REF}$ LS by a factor 26$\times$ in 65~nm (resp. 12.5$\times$ in 22~nm) and comes for free in the proposed reference as $V_{G8}$ is readily available from the body$\:$/$\:$back-gate bias generator.

\vspace{-0.25cm}
\subsection{Parasitic Diode Leakage Mitigation}
\label{subsec:4B_parasitic_diode_leakage_mitigation}
\begin{figure}[!t]
	\centering
	\includegraphics[width=.45\textwidth]{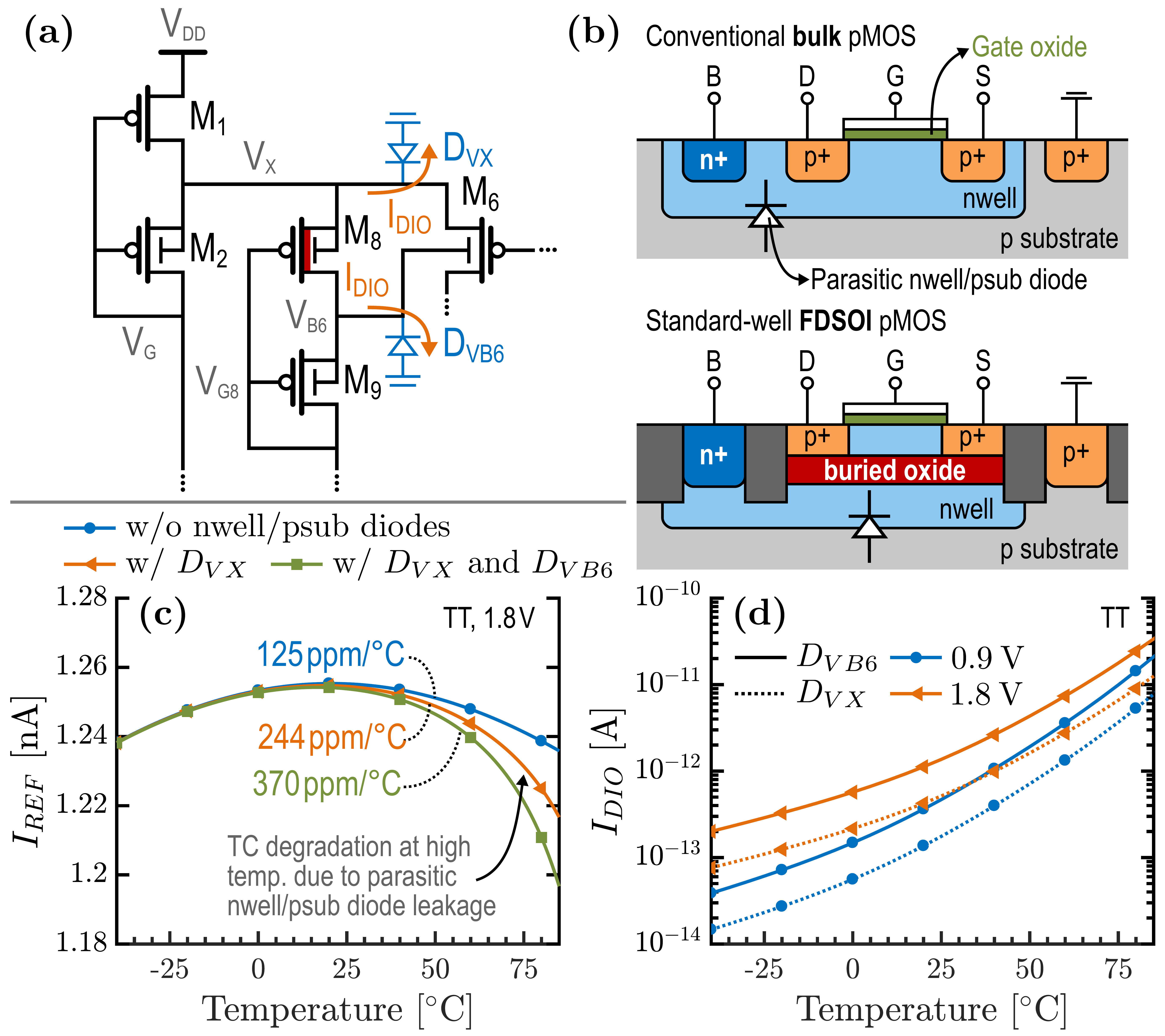}
	\caption{$I_{REF}$ TC is degraded by the leakage of parasitic nwell/psub diodes. (a) Schematic of the SCM and body$\:$/$\:$back-gate bias generator with parasitic diodes. (b) Cross-section of pMOS devices in conventional bulk and standard-well FDSOI technologies. (c) Temperature dependence of $I_{REF}$ in TT 1.8~V, based on pre- and post-layout simulations in 22~nm. (d) Diode leakage as a function of temperature in 22~nm, for reverse voltages of 0.9 and 1.8~V.}
	\label{fig:12_parasitic_nwell_psub_diode}
\end{figure}
\begin{figure}[!t]
	\centering
	\includegraphics[width=.45\textwidth]{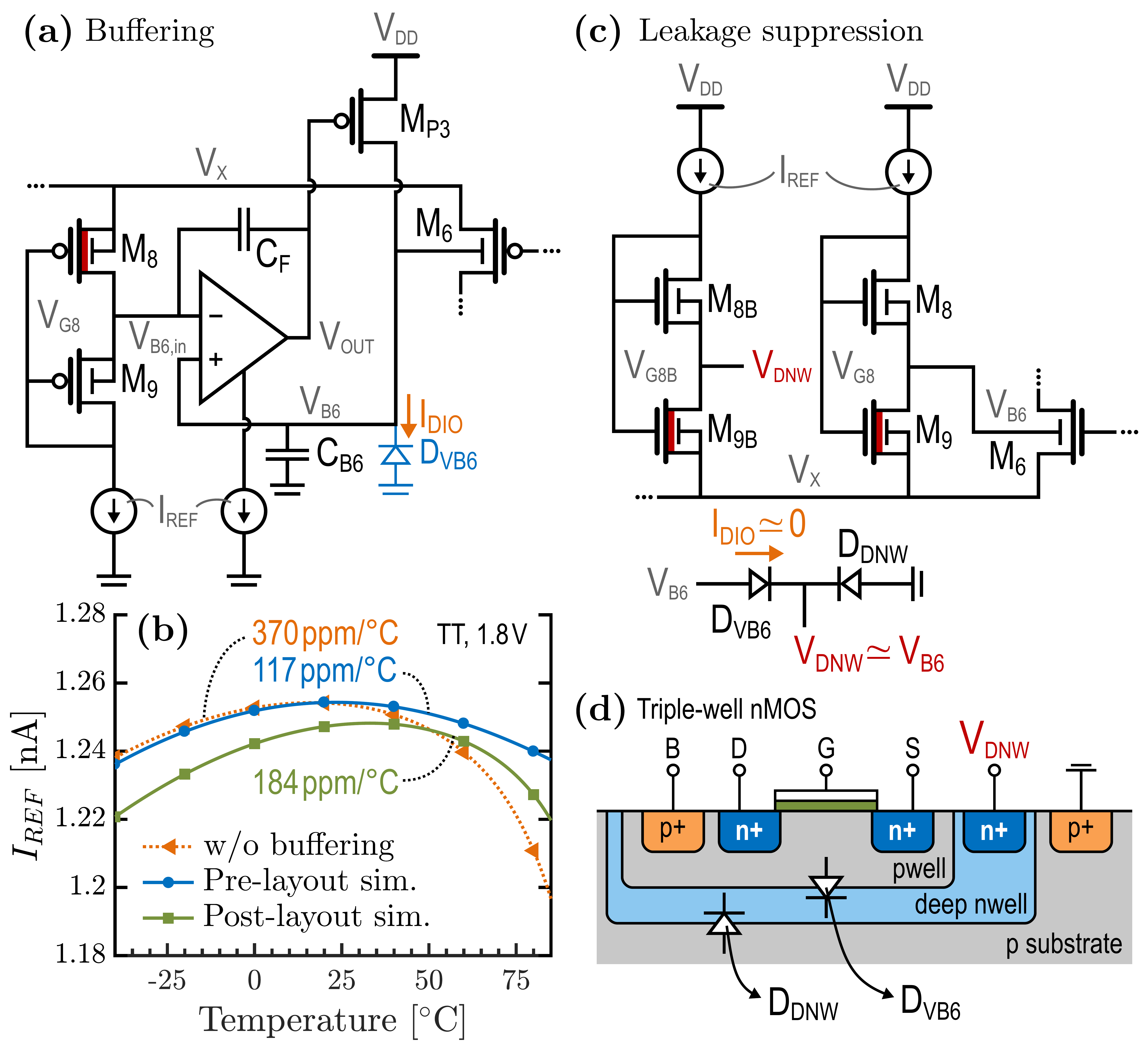}
	\caption{Schematic of the body$\:$/$\:$back-gate bias generator (a) with a buffering stage implemented as an LDO, or (c) with a leakage suppression mechanism, obtained by implementing $M_6$ with a triple-well nMOS device whose deep nwell is biased by a body bias generator replica. (b) Temperature dependence of $I_{REF}$ in pre- and post-layout simulations, with a fine-tuning of the SCM sizes in TT 1.8~V. (d) Cross-section of triple-well nMOS devices.}
	\label{fig:13_vb6_buffering}
\end{figure}
\begin{figure}[!t]
	\centering
	\includegraphics[width=.422\textwidth]{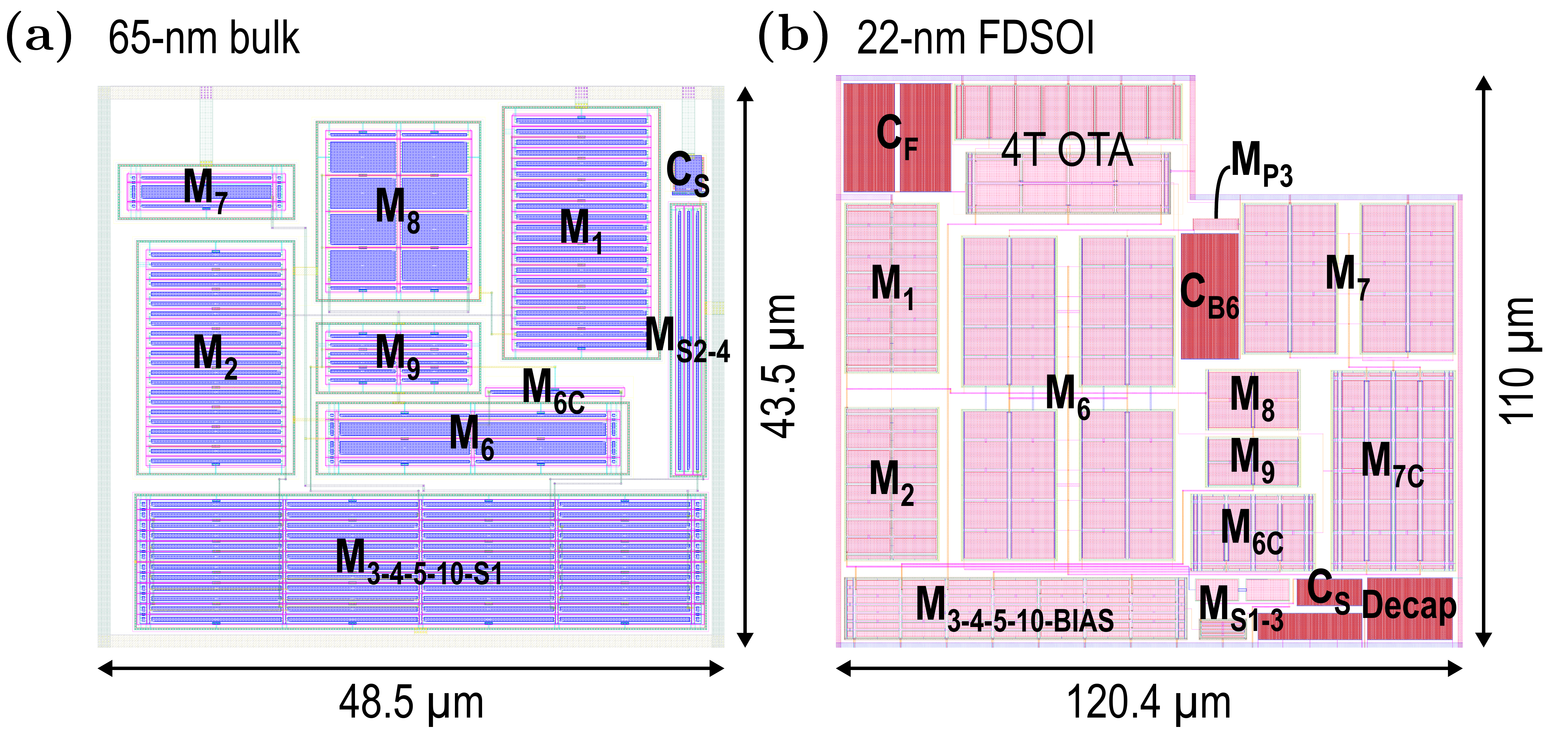}
	\caption{Layout of the reference in (a) \mbox{65-nm} bulk and (b) \mbox{22-nm} FDSOI.}
	\label{fig:15_layout}
\end{figure}
In the case of a pMOS-based SCM [Fig.~\ref{fig:12_parasitic_nwell_psub_diode}(a)] and at high temperature, $I_{REF}$ TC is degraded by leakage currents flowing from $V_X$ and $V_{B6}$ to ground through parasitic nwell/psub diodes. The origin of these diodes is illustrated by the cross-section of pMOS devices in Fig.~\ref{fig:12_parasitic_nwell_psub_diode}(b). This issue does not occur in 65~nm due to the lower reverse voltage applied to the diodes and the limited nwell area required by $M_{6-6C-9}$, as smaller devices can be used because of the low $S_{I_{REF}}$. Therefore, we only discuss the \mbox{22-nm} design in what follows. Next, Fig.~\ref{fig:12_parasitic_nwell_psub_diode}(c) shows that in 22~nm, the pre-layout simulation in TT 1.8~V features a \mbox{125-ppm/$^\circ$C} TC, while post-layout simulations with either diode $D_{VX}$ or both diodes $D_{VX}$ and $D_{VB6}$ result in a degraded $I_{REF}$ TC of 244 and 370~ppm/$^\circ$C, respectively. This degradation arises from the exponential increase of diode leakage with temperature [Fig.~\ref{fig:12_parasitic_nwell_psub_diode}(d)], causing a deviation from the pre-layout behavior above 50$^\circ$C. $D_{VB6}$'s leakage attains 21 to 33.8~pA at 85$^\circ$C and mostly has an indirect impact on $I_{REF}$ by inducing a change in $V_{B6}$, passed on $I_{REF}$ through $\Delta V_T$. Furthermore, $D_{VX}$'s leakage at 85$^\circ$C lies between 7.8 and 12.5~pA and has a direct impact on $I_{REF}$ by drawing some current away from the SCM and $\beta$-multiplier. However, it should be noted that this impact can be alleviated by connecting the body of $M_2$ to $V_{DD}$ while accounting for this change in the sizing of the SCM.\\
\indent In Fig.~\ref{fig:13_vb6_buffering}, two solutions are proposed to combat the adverse effect of parasitic diode leakage. First, a low-dropout (LDO) regulator copes with $D_{VB6}$'s leakage by buffering node $V_{B6,\mathrm{in}}$ [Fig.~\ref{fig:13_vb6_buffering}(a)]. It consists of four main components: (i) a self-biased 4T operational transconductance amplifier (OTA) with an nMOS differential pair, (ii) a regulation pMOS $M_{P3}$ delivering the leakage current of $D_{VB6}$, (iii) a capacitor $C_{B6}$ ensuring the stability of the LDO, and (iv) a Miller capacitor $C_F$ ensuring the stability of the whole reference by adding a dominant pole $f_{pd} = (g_{m8}+g_{d8})/(2\pi C_F A_{v,OTA})$ at node $V_{B6,\mathrm{in}}$, with $A_{v,OTA}$ the differential voltage gain of the OTA in DC conditions. Fig.~\ref{fig:13_vb6_buffering}(b) reveals that a \mbox{117-ppm/$^\circ$C} $I_{REF}$ TC is obtained pre-layout with this additional buffering stage. With a fine-tuning of the SCM sizes to alleviate the impact of $D_{VX}$'s residual leakage, an increase of the TC to 184~ppm/$^\circ$C is observed post-layout, albeit the degradation is lessened compared to the 370~ppm/$^\circ$C without buffering. At last, the buffering stage entails power and area overheads corresponding to 14.3~$\%$ (0.83~nW) and 14.7~$\%$ (1940~$\mu$m$^2$) of the total power and area usage.\\
\indent Second, in the case of an nMOS-based SCM, leakage suppression could be achieved by biasing the deep nwell with a voltage $V_{DNW} \simeq V_{B6}$. This technique is inspired from ultra-low-power voltage references \cite{Fassio_2021}, and forces a voltage drop close to zero across the parasitic pwell/deep-nwell diode $D_{VB6}$ [Fig.~\ref{fig:13_vb6_buffering}(c)], thus suppressing the leakage from $V_{B6}$. Voltage $V_{DNW}$ is easily generated by a replica of the low-power voltage reference which provides the leakage of $D_{DNW}$. The origin of diodes $D_{VB6}$ and $D_{DNW}$ is illustrated by the cross-section of triple-well devices in Fig.~\ref{fig:13_vb6_buffering}(d).

\vspace{-0.25cm}
\subsection{Final Implementation Performance}
\label{subsec:4C_final_implementation_performance}
\begin{figure}[!t]
	\centering
	\includegraphics[width=.425\textwidth]{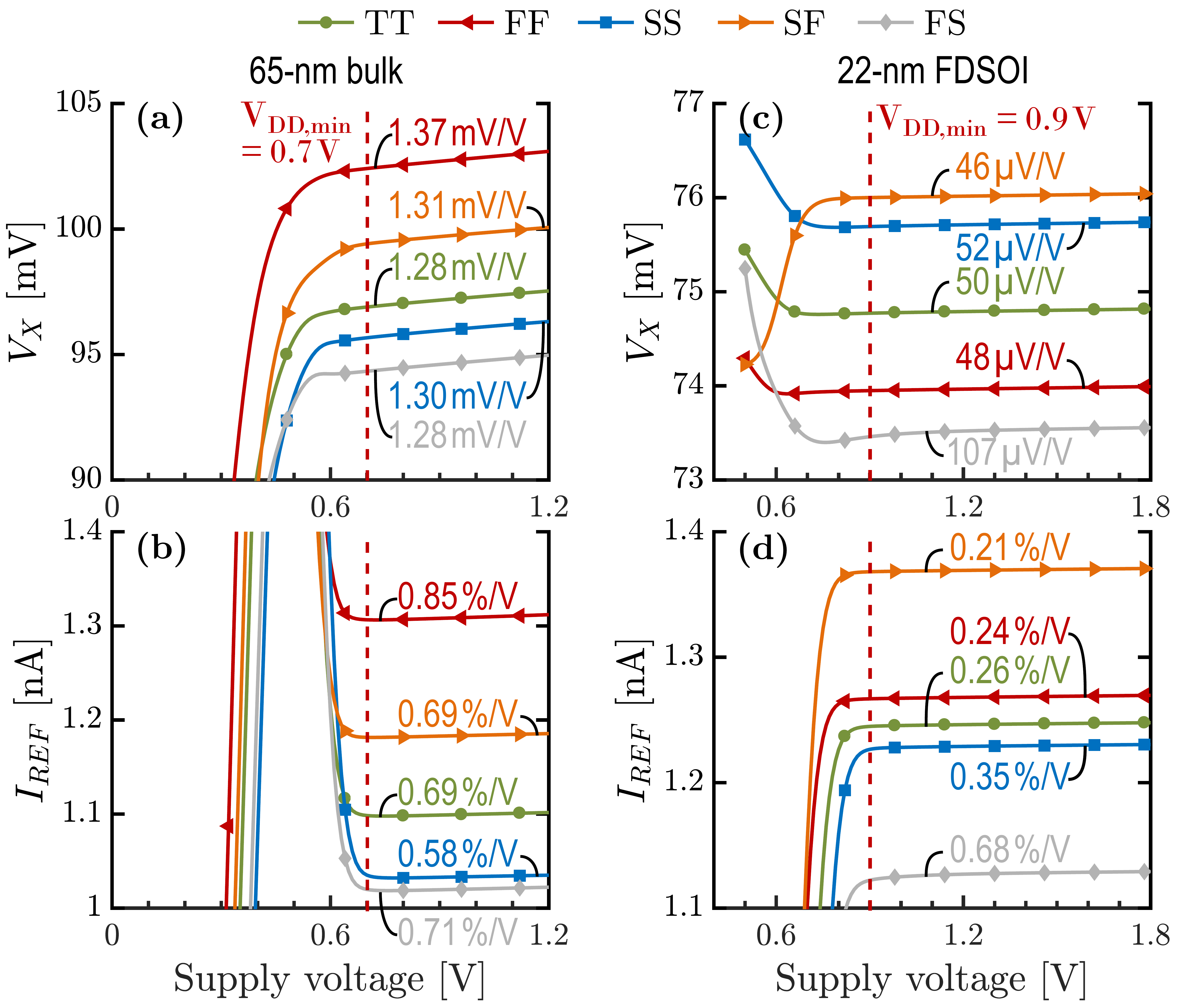}
	\caption{Post-layout simulations of the supply voltage dependence of $V_X$ and $I_{REF}$ in all process corners and at 25$^\circ$C, in \mbox{65-nm} bulk [(a) and (b)] and in \mbox{22-nm} FDSOI [(c) and (d)].}
	\label{fig:16_sim_iref_vs_vdd}
\end{figure}

\begin{figure}[!t]
	\centering
	\includegraphics[width=.425\textwidth]{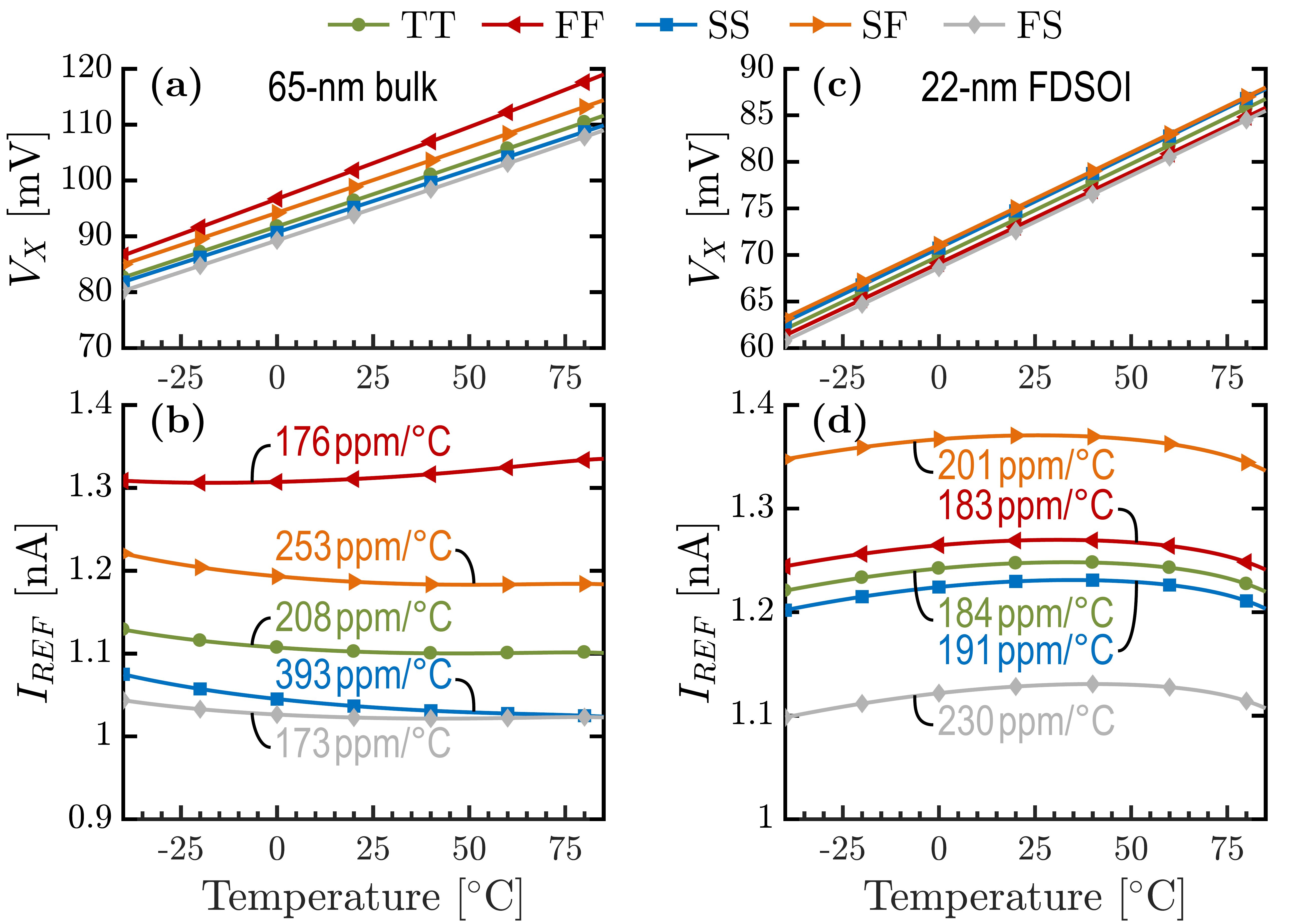}
	\caption{Post-layout simulations of the temperature dependence of $V_X$ and $I_{REF}$ in all process corners, at 1.2~V in \mbox{65-nm} bulk [(a) and (b)] and at 1.8~V in \mbox{22-nm} FDSOI [(c) and (d)].}
	\label{fig:17_sim_iref_vs_T}
\end{figure}

\begin{figure}[!t]
	\centering
	\includegraphics[width=.45\textwidth]{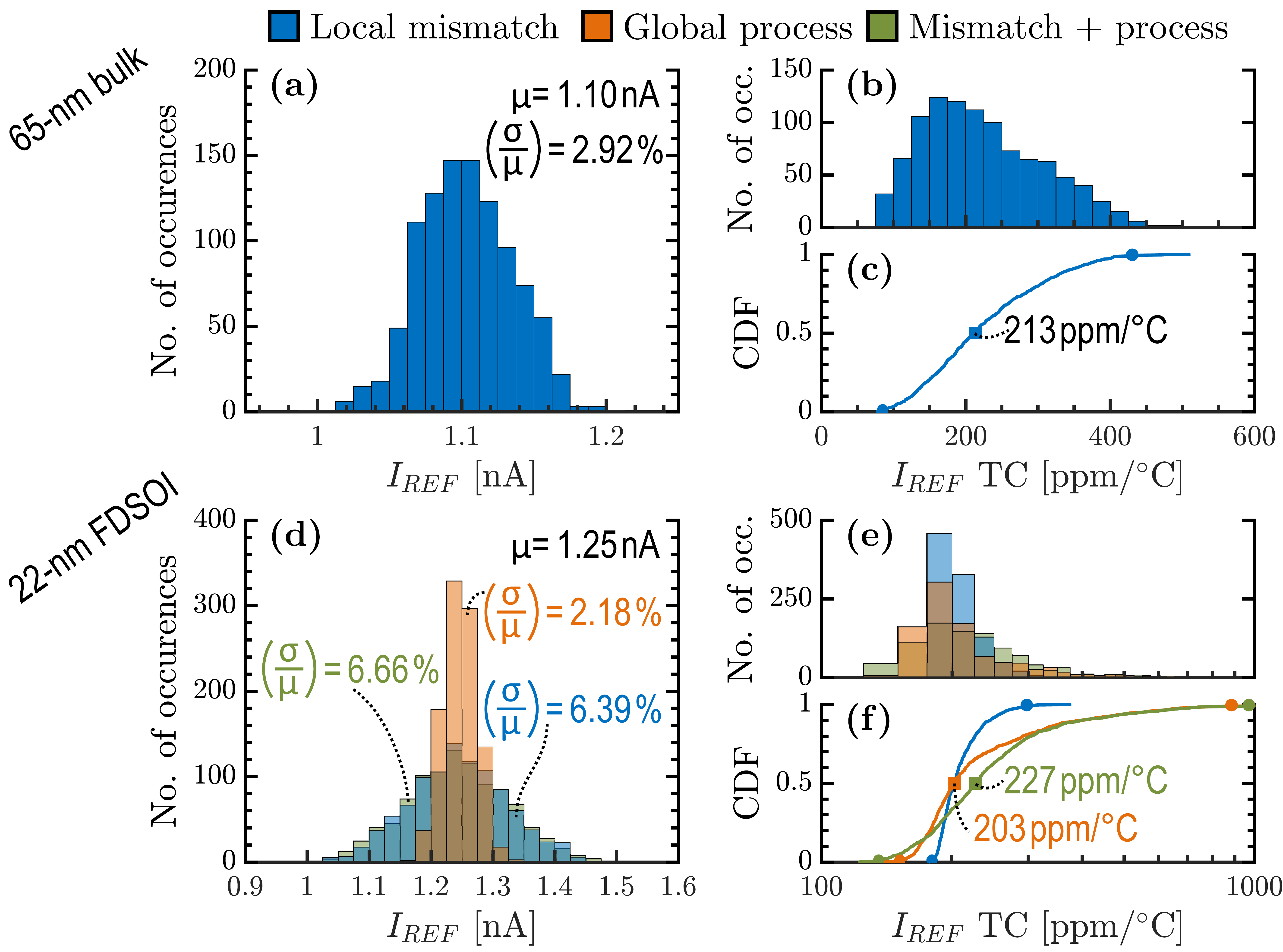}
	\caption{For 10$^3$ post-layout MC simulations in TT, histograms of $I_{REF}$ at 25$^\circ$C, and $I_{REF}$ TC from \mbox{-40} to 85$^\circ$C, at 1.2~V in \mbox{65-nm} bulk [(a) to (c)] and 1.8~V in \mbox{22-nm} FDSOI [(d) to (f)]. (d) to (f) correspond to simulations of local mismatch, global process variations, and their combined effects.}
	\label{fig:18_sim_iref_vs_T_mc}
\end{figure}

\begin{figure}[!t]
	\centering
	\includegraphics[width=.4\textwidth]{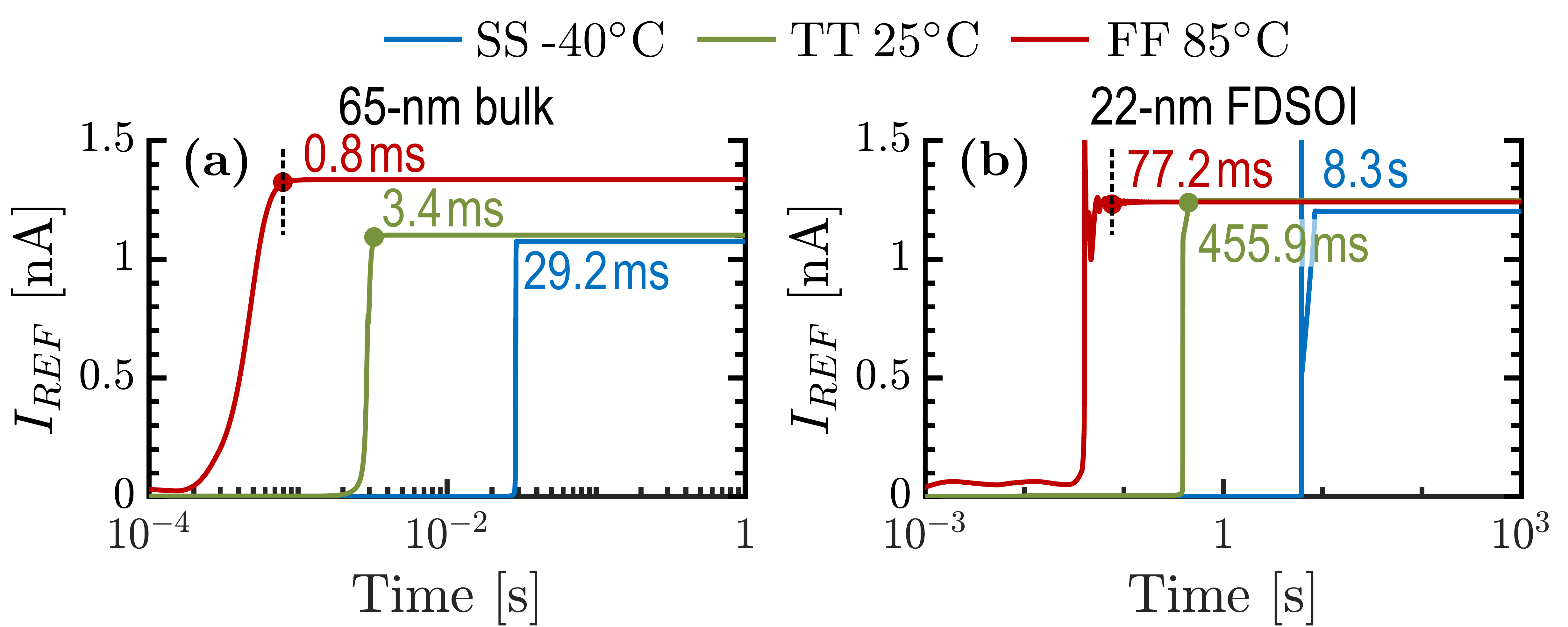}
	\caption{Post-layout startup waveforms in the fastest (FF 85$^\circ$C), typical (TT 25$^\circ$C), and slowest (SS \mbox{-40$^\circ$C}) corners, (a) at 1.2~V in \mbox{65-nm} bulk and (b) 1.8~V in \mbox{22-nm} FDSOI in the three process and temperature corners, using a voltage source with a rise time of 100~$\mu$s.}
	\label{fig:19_sim_iref_startup}
\end{figure}

The complete schematic of the proposed reference is disclosed in Fig.~\ref{fig:14_final_implementation_schematic} for the 65- and \mbox{22-nm} designs, and features three additions compared to Fig.~\ref{fig:6_pmos_implementation_schematic}. Firstly, a dynamic startup circuit \cite{CamachoGaleano_2005, CamachoGaleano_2008}, to make sure the reference is not stuck around its zero-$I_{REF}$ stable operation point. For applications relying on a low-slew-rate supply voltage, a suitable startup circuit can be obtained by replacing $C_S$ by a diode-connected pMOS or by relying on the conventional circuit in \cite{Liu_1998}, at the cost of an increased power consumption. Second, a buffering stage for $V_{B6}$ in 22~nm. Lastly, a low-voltage cascode to improve LS, which is either implemented with a single transistor in 65~nm, or in full in 22~nm. Our analysis in Section~\ref{subsec:4A_line_sensitivity_improvement} nevertheless suggests that the \mbox{22-nm} cascode could be simplified to a single transistor without loss of performance. Furthermore, a pMOS-based SCM has been selected to avoid the minor area penalty and slightly more complex body$\:$/$\:$back-gate connections entailed by the use of triple-well devices, albeit a design relying on an nMOS-based SCM is totally feasible. Then, the layout is presented in Fig.~\ref{fig:15_layout}, and transistor sizes are summarized in Table~\ref{table:final_implementation_sizes}, with core (resp. I/O) LVT devices used by default in 65~nm (resp. 22~nm). Most transistors present relatively small aspect ratios, as they are biased with a current in the nA range lower than their specific sheet current $I_{SQ}$ and, for some of them, are operated in moderate inversion, thus reducing their aspect ratio even further. This effect is more pronounced in the \mbox{65-nm} design because the body$\:$/$\:$back-gate bias generator produces a \mbox{28-mV} $\Delta V_T$, as opposed to 17~mV in 22~nm, which pushes $M_{1-2}$ closer to strong inversion.\\
\indent Next, we discuss the performance of the proposed reference. First, Fig.~\ref{fig:16_sim_iref_vs_vdd} illustrates the dependence to supply voltage, with the minimum supply voltage given by
\begin{IEEEeqnarray}{RCL}
	V_{DD,\mathrm{min}} & = & V_{DS,\mathrm{sat}} + \max\big(V_{SG1}, V_{SG8} + V_{SG1} - V_{SG2},\IEEEnonumber\\
	& & V_{GS4} + V_{SD6C,\mathrm{sat}} + V_{SG1} - V_{SG2}, V_{SG7}\big)\label{eq:vdd_min}\textrm{,}
\end{IEEEeqnarray}
in which each expression inside the max function corresponds to one of the branches of the schematic in Fig.~\ref{fig:14_final_implementation_schematic}(a). Second, Fig.~\ref{fig:17_sim_iref_vs_T} displays the temperature dependence of $V_X$ and $I_{REF}$. Third, Fig.~\ref{fig:18_sim_iref_vs_T_mc} studies the variability of $I_{REF}$ and its TC for 10$^3$ Monte-Carlo (MC) simulations. Note that $I_{REF}$ variability can be estimated from $\sigma_{V_X}$ at first order using
\begin{table}[!t]
\centering
\caption{Sizing of the proposed nA-range CWT reference.}
\label{table:final_implementation_sizes}
\begin{scriptsize}
\begin{tabular}{lcccccc}
	\toprule
	& \multicolumn{3}{c}{TSMC \mbox{65-nm} bulk} & \multicolumn{3}{c}{GF \mbox{22-nm} FDSOI}\\
	\cmidrule(l){2-4} \cmidrule(l){5-7}
	& $W$ [$\mu$m] & $L$ [$\mu$m] & $i_f$ & $W$ [$\mu$m] & $L$ [$\mu$m] & $i_f$\\
	\midrule
	$M_1$ & 0.365 & 20$\times$10 & 52.46 & 2.39 & 20$\times$8 & 5.17\\
	$M_2$ & 0.3 & 20$\times$10 & 22.09 & 2.74 & 16$\times$8 & 1.10\\
	$M_3$ & 3$\times$0.35 & 5$\times$10 & 0.69 & 1 & 5$\times$8 & 0.25\\
	$M_4$ & 0.35 & 5$\times$10 & 0.69 & 1 & 5$\times$8 & 0.25\\
	$M_5$ & 2$\times$0.35 & 5$\times$10 & 0.69 & 3$\times$1 & 5$\times$8 & 0.25\\
	$M_6$ & 4$\times$1 & 10 & 0.03 & 20$\times$5 & 2$\times$8 & 0.003\\
	$M_7$ & 1 & 10 & 0.03 & 10$\times$5 & 2$\times$8 & 0.003\\
	$M_{6C}$ & 0.22 & 10 & - & 8$\times$5 & 4 & -\\
	$M_{7C}$ & - & - & - & 24$\times$5 & 4 & -\\
	$M_8$ & 2$\times$2.34 & 4$\times$5 & - & 4$\times$5 & 8 & -\\
	$M_9$ & 2$\times$0.22 & 4$\times$5 & - & 4$\times$4 & 8 & -\\
	$M_{10}$ & 0.35 & 5$\times$10 & 0.69 & 1 & 5$\times$8 & 0.25\\
	$M_{S1}$ & 0.35 & 5$\times$10 & 0.69 & 0.16 & 5$\times$8 & 0.25\\
	$M_{S2-S4}$ & 0.2 & 20 & - & 4 & 8 & -\\
	$C_{S}$ & 2.5 & 2 & - & 12.5 & 5 & -\\
	\bottomrule
\end{tabular}
\end{scriptsize}
\end{table}
\begin{equation}
	\left(\sigma/\mu\right)_{I_{REF}} = S_{I_{REF}} \sigma_{V_X}\label{eq:var_iref}\textrm{.}
\end{equation}
Finally, Fig.~\ref{fig:19_sim_iref_startup} depicts the startup waveforms in nominal and extreme corners for a \mbox{100-$\mu$s} \mbox{rise-time} voltage source.\\
\indent In 65~nm, the layout [Fig.~\ref{fig:15_layout}(a)] is dominated by the SCM which occupies 23.8~$\%$ of the \mbox{0.0021-mm$^2$} area, followed by the current mirrors with 22.8~$\%$. In TT, the LS of $V_X$ is equal to 1.28~mV/V from 0.7 to 1.2~V [Fig.~\ref{fig:16_sim_iref_vs_vdd}(a)] and yields a 0.69-$\%$V LS of $I_{REF}$ [Fig.~\ref{fig:16_sim_iref_vs_vdd}(b)]. Process variations have negligible impact on the LS but change $I_{REF}$ by +19.09~$\%$ in FF and -7.20~$\%$ in SF. Fig.~\ref{fig:17_sim_iref_vs_T}(a) highlights the slightly different TC of $V_X$ due to process variations of the subthreshold slope factor. These differences result in $I_{REF}$ TC discrepancies, with a typical and worst-case values of 208 and 393~ppm/$^\circ$C. Regarding variability, only mismatch can be simulated in 65~nm as no statistical models for global process variations are available in the process design kit (PDK) used in this work. Fig.~\ref{fig:18_sim_iref_vs_T_mc}(a) exhibits a \mbox{2.92-$\%$} ($\sigma/\mu$) lower than the 3.74~$\%$ inferred from (\ref{eq:var_iref}) and the \mbox{1.37-mV} $\sigma_{V_X}$, concurring with the hypothesis of an $S_{I_{REF}}$ overestimation. Regarding TC, its median and 99$^\mathrm{th}$ percentile are 213 and 431~ppm/$^\circ$C [Figs.~\ref{fig:18_sim_iref_vs_T_mc}(b) and (c)]. Finally, the typical and worst-case startup times $t_s$ are 3.4 and 29.2~ms [Fig.~\ref{fig:19_sim_iref_startup}(a)].\\
\indent In 22~nm, the layout [Fig.~\ref{fig:15_layout}(b)] is dominated by $M_{6-7}$, which need to be large to minimize $\sigma_{V_X}$ and thus account for 28.7~$\%$ of the \mbox{0.0132-mm$^2$} area. Then, regarding supply voltage dependence, the slender process variations impacting $V_X$ [Fig.~\ref{fig:16_sim_iref_vs_vdd}(c)] stem from $V_{B6}$ through $\Delta V_T$. However, $I_{REF}$ is mainly impacted by $I_{SQ}$ variations [Fig.~\ref{fig:16_sim_iref_vs_vdd}(d)], and changes by at most +9.85~$\%$ in SF and -9.51~$\%$ in FS. From 0.9 to 1.8~V, the LS of $V_X$ is worth 50~$\mu$V/V in TT, and translates to a \mbox{0.26-$\%$/V} LS of $I_{REF}$. Furthermore, LS is relatively constant across process corners except in FS, for which the \mbox{0.68-$\%$/V} value can be explained by the distortion close to 0.9~V. It requires a higher $V_{DD}$ to cope with the increased pMOS $V_T$. Then, $V_X$ has the same TC across process corners [Fig.~\ref{fig:17_sim_iref_vs_T}(c)], leading to similar $I_{REF}$ TCs in Fig.~\ref{fig:17_sim_iref_vs_T}(d). The typical and worst-case TCs are 184 and 230~ppm/$^\circ$C, the latter being partially explained by the lower $I_{REF}$ value in this process corner, as $I_{REF}$ TC is a relative metric. As far as variability is concerned, in 22~nm, we can simulate the impact of mismatch, process variations, and their combined effects. On the one hand, mismatch has a strong impact on $I_{REF}$ ($\sigma/\mu$) due to the large $S_{I_{REF}}$. The simulated \mbox{0.87-mV} $\sigma_{V_X}$ leads to a \mbox{4.92-$\%$} ($\sigma/\mu$) prediction, which falls short of the \mbox{6.39-$\%$} simulation [Fig.~\ref{fig:18_sim_iref_vs_T_mc}(d)] due to additional mismatch in the SCM and the mirrors. Besides, TC variability is relatively limited, with a 203- and \mbox{299-ppm/$^\circ$C} median and 99$^\mathrm{th}$ percentile [Figs.~\ref{fig:18_sim_iref_vs_T_mc}(e) and (f)]. On the other hand, process variations have a restricted impact on $I_{REF}$, with a \mbox{2.18-$\%$} ($\sigma/\mu$), but deteriorate $I_{REF}$ TC, with a 99$^\mathrm{th}$ percentile shifted to 887~ppm/$^\circ$C due to a malfunction of the buffer for extreme process realizations causing $I_{REF}$ to drop at high temperature. Unsurprisingly, the combined effects yield a \mbox{6.66-$\%$} ($\sigma/\mu$) for $I_{REF}$, together with a 227- and \mbox{972-ppm/$^\circ$C} median and 99$^\mathrm{th}$ percentile for $I_{REF}$ TC. Lastly, the typical and worst-case $t_s$ are 456~ms and 8.3~s [Fig.~\ref{fig:19_sim_iref_startup}(b)], because of the parasitic caps of large transistors. While the worst-case \mbox{8.3-s} $t_s$ could be critical, most applications require the reference to operate down to \mbox{-40$^\circ$C} but not to start in the extreme SS \mbox{-40$^\circ$C} corner, as nA-range current references are often always-on circuits. Nevertheless, this issue can be solved by widening $M_{S2-S3}$ in Fig.~\ref{fig:14_final_implementation_schematic}(b), at the expense of a TC degradation in FF 85$^\circ$C.
\begin{figure}[!t]
	\centering
	\includegraphics[width=.469\textwidth]{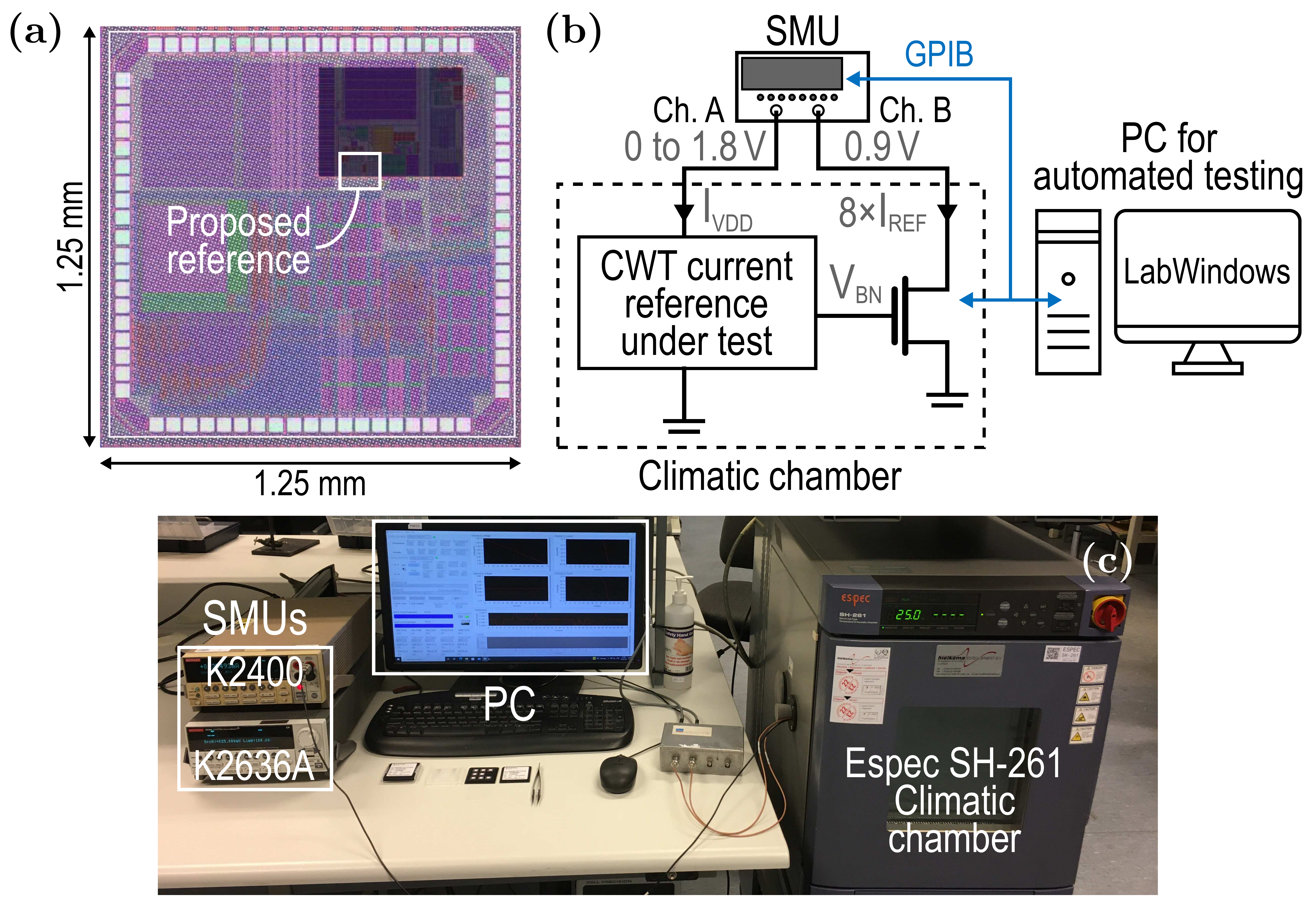}
	\caption{(a) Chip microphotograph in GF \mbox{22-nm} FDSOI, with the proposed reference highlighted in white. (b) Conceptual and (c) actual measurement testbench for supply voltage and temperature dependence characterization.}
	\label{fig:20_meas_setup}
\end{figure}

\vspace{-0.25cm}
\section{Measurement Results}
\label{sec:5_measurement_results}
\begin{figure}[!t]
	\centering
	\includegraphics[width=.4\textwidth]{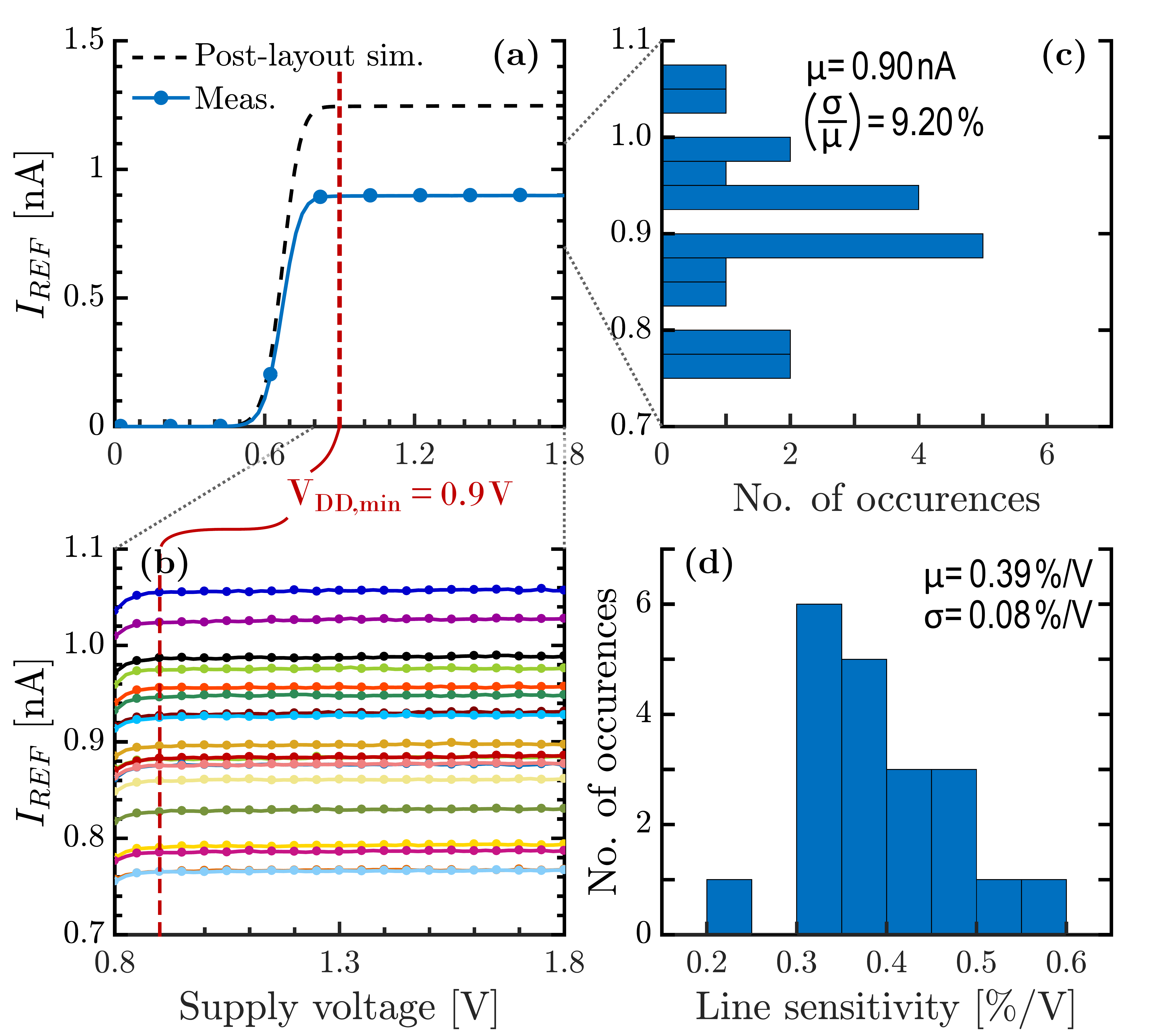}
	\caption{(a) Post-layout-simulated and measured average $I_{REF}$, with (b) details of the 20 dies. Measured histograms of (c) $I_{REF}$ at 1.8~V and (d) LS from 0.9 to 1.8~V. All figures are obtained at a temperature of 25$^\circ$C.}
	\label{fig:21_meas_iref_vs_vdd}
\end{figure}

\begin{figure}[!t]
	\centering
	\includegraphics[width=.45\textwidth]{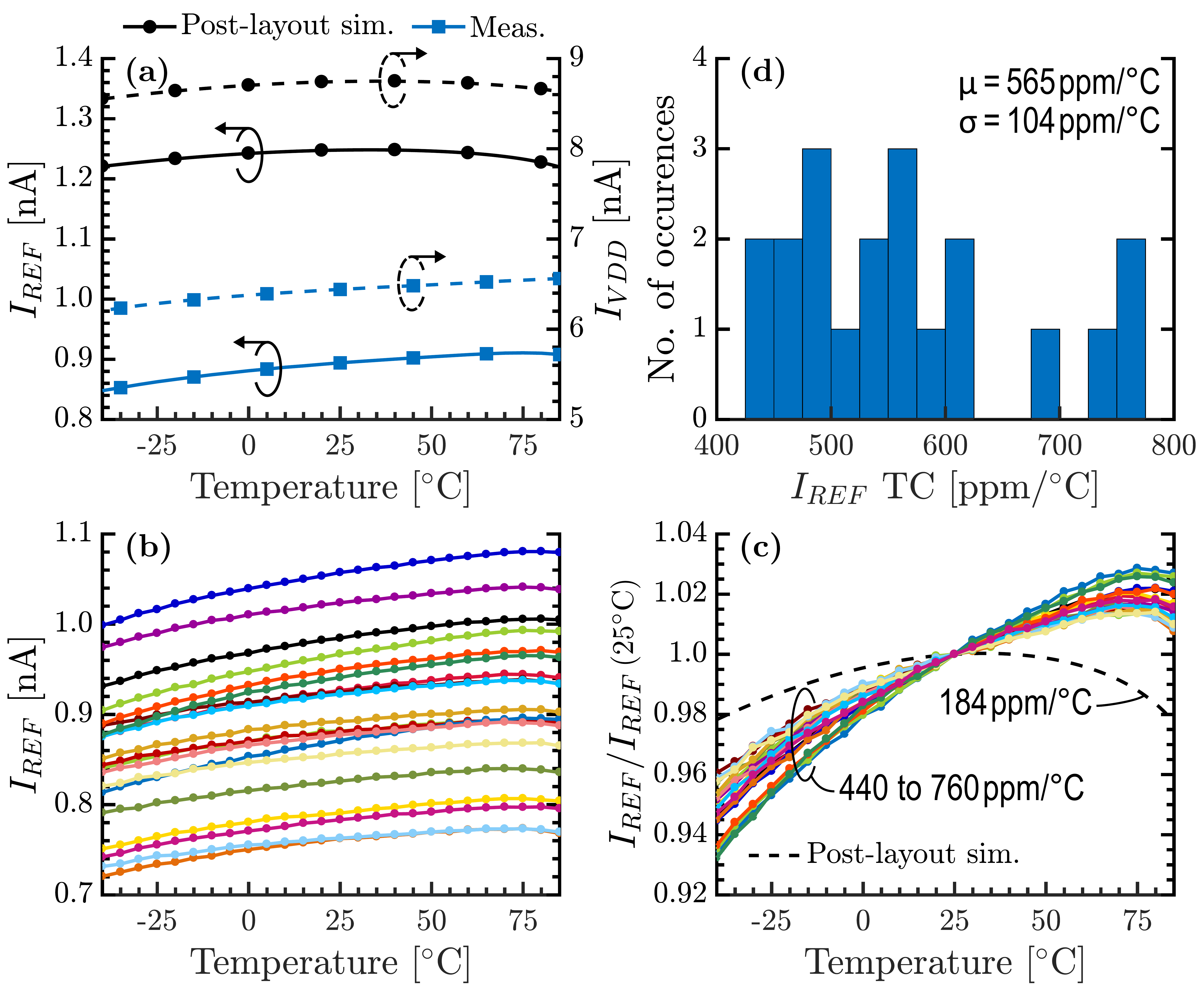}
	\caption{(a) Post-layout-simulated and measured temperature dependence of the average $I_{REF}$ and $I_{VDD}$. Measured temperature dependence of $I_{REF}$ (b) without and (c) with normalization by the value at 25$^\circ$C, for the 20 dies. (d) Measured histogram of $I_{REF}$ TC from \mbox{-40} to 85$^\circ$C. All figures are obtained for a supply voltage of 1.8~V.}
	\label{fig:22_meas_iref_vs_T}
\end{figure}

In this section, we describe the testbench used to characterize the proposed current reference, fabricated in GF 22-nm FDSOI as part of a \mbox{1.56-mm$^2$} microcontroller unit for biomedical applications codenamed ICare [Fig.~\ref{fig:20_meas_setup}(a)]. Then, we present the LS and TC measurements across the 20 dies in Figs.~\ref{fig:21_meas_iref_vs_vdd} and \ref{fig:22_meas_iref_vs_T}, before discussing the discrepancies with respect to post-layout simulations based on Figs.~\ref{fig:23_sim_meas_iref_cross_process} and \ref{fig:24_sim_meas_iref_vs_T}.

\vspace{-0.25cm}
\subsection{Measurement Testbench}
Conceptual and actual views of the testbench are shown in Figs.~\ref{fig:20_meas_setup}(b) and (c). Supply voltage dependence is characterized using a two-channel Keithley K2636A source measurement unit (SMU), providing the supply voltage to the reference on channel A, and a \mbox{0.9-V} $V_{DS}$ to the output nMOS transistor on channel B. The supply current $I_{VDD}$ and the output current, equal to $8\times I_{REF}$, are measured by the K2636A, while a K2400 supplies the I/O voltage [Fig.~\ref{fig:20_meas_setup}(c)]. Supply voltage is then swept from 0 to 1.8~V in \mbox{25-mV} steps. Then, temperature dependence is characterized using an Espec SH-261 climatic chamber which sweeps temperature from \mbox{-40} to 85$^\circ$C in 5$^\circ$C steps. Both equipments are connected in GPIB to a PC running a testing routine in LabWindows. Lastly, LS and TC are computed across the 20 dies according to (\ref{eq:LS}) and (\ref{eq:TC}).

\vspace{-0.25cm}
\subsection{Supply Voltage and Temperature Dependence}
\begin{figure}[!t]
	\centering
	\includegraphics[width=.45\textwidth]{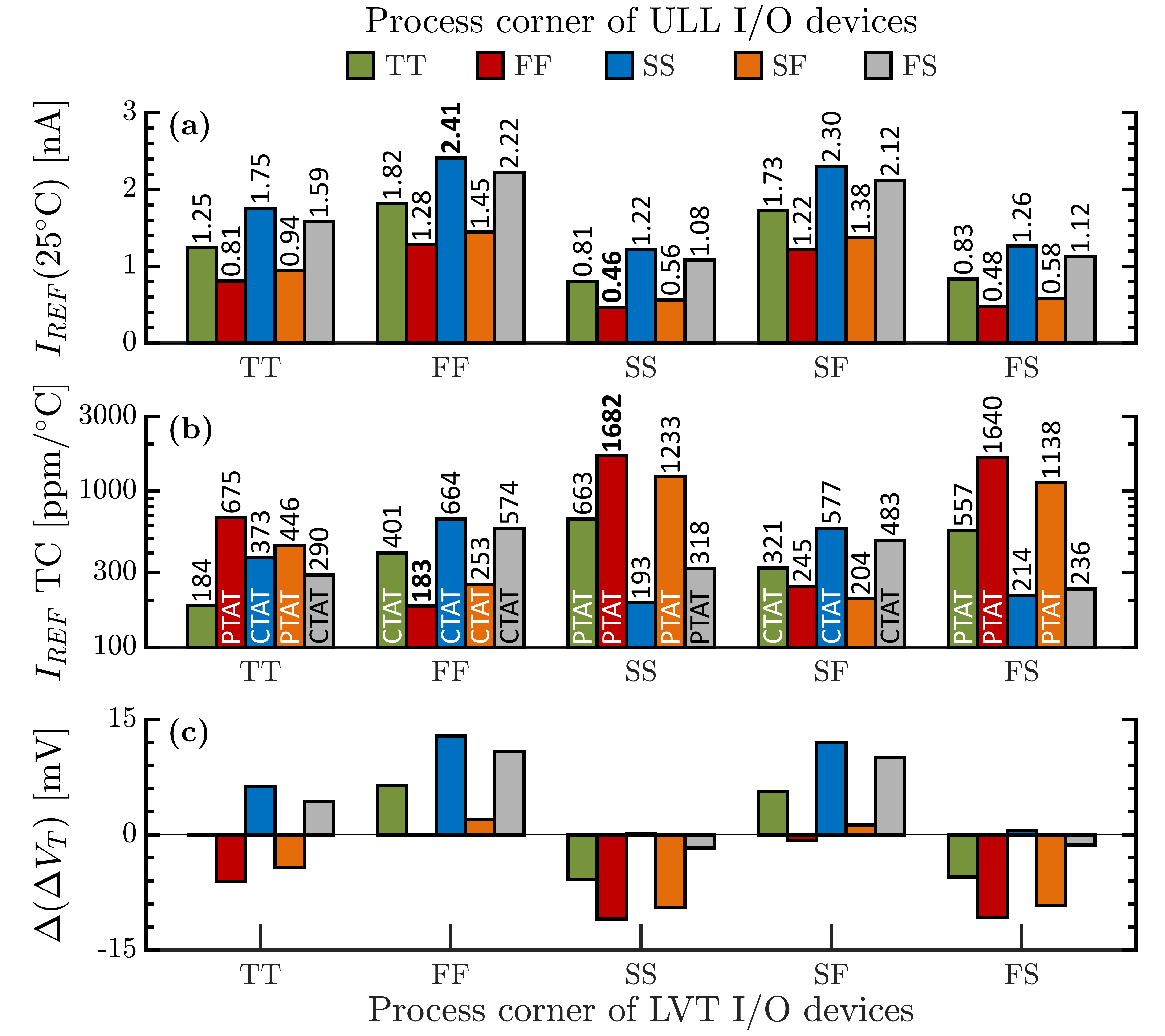}
	\caption{Skewed process corners of ULL and LVT I/O devices have an impact on (a) $I_{REF}$ at 25$^\circ$C and (b) $I_{REF}$ TC from -40 to 85$^\circ$C, due to (c) changes in the CWT offset $\Delta V_{T}$ with respect to its nominal value. All results are obtained for a supply voltage of 1.8~V and a typical process for parasitic nwell/psub diodes.}
	\label{fig:23_sim_meas_iref_cross_process}
\end{figure}

\begin{figure}[!t]
	\centering
	\includegraphics[width=.45\textwidth]{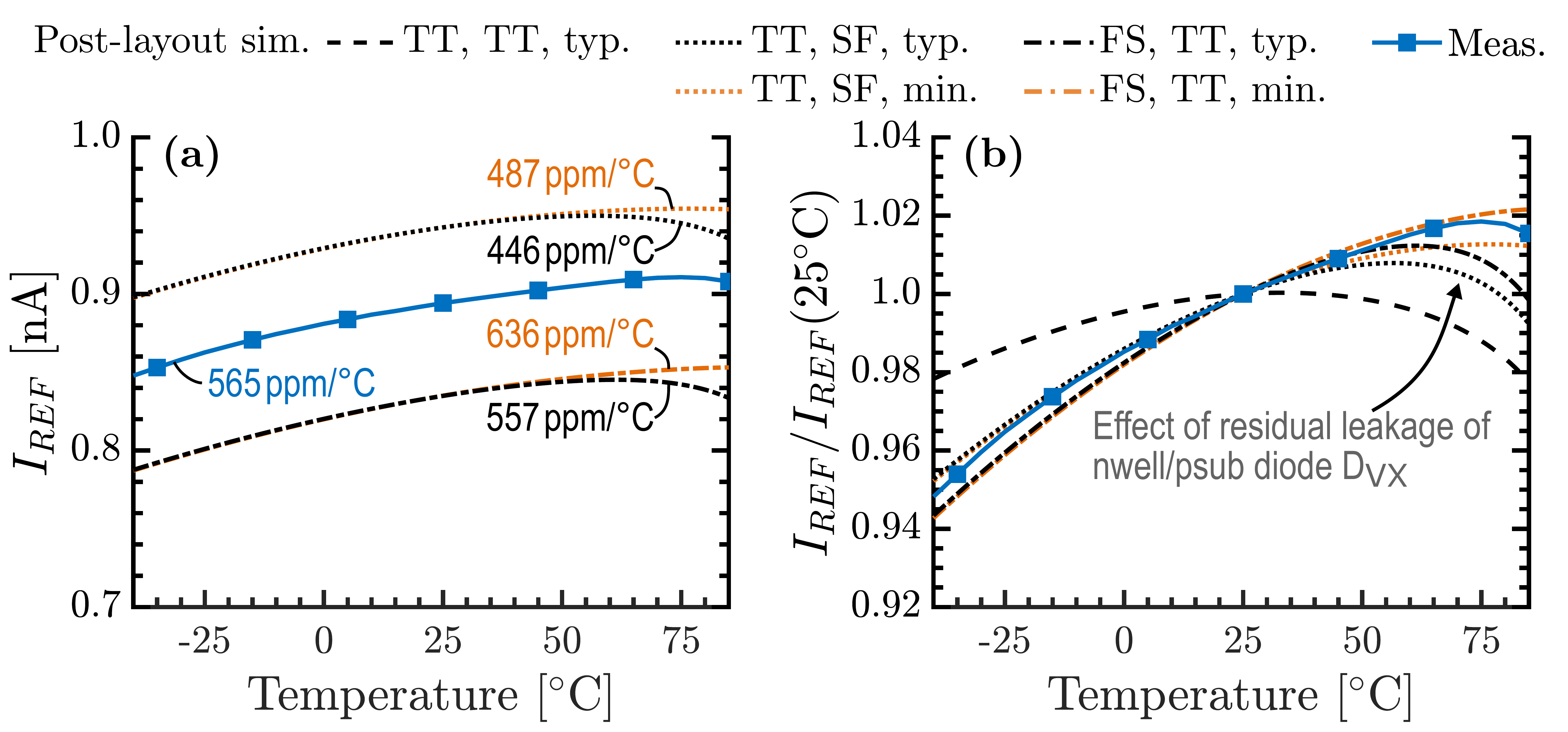}
	\caption{Post-layout-simulated and measured temperature dependence of $I_{REF}$ at 1.8~V (a) without and (b) with normalization by the value at 25$^\circ$C. Post-layout simulations correspond to skewed process corners for LVT I/O devices, ULL I/O devices, and parasitic nwell/psub diodes.}
	\label{fig:24_sim_meas_iref_vs_T}
\end{figure}

Firstly, Fig.~\ref{fig:21_meas_iref_vs_vdd} quantifies the LS and variability of $I_{REF}$ obtained in measurement at 25$^\circ$C. Fig.~\ref{fig:21_meas_iref_vs_vdd}(a) describes the supply voltage dependence of $I_{REF}$ for post-layout simulations in TT and for the measurements average. The measured $V_{DD,\mathrm{min}}$ remains about 0.9~V as expected from simulation, but $I_{REF}$ shrinks from 1.25 to 0.9~nA. This difference cannot be explained by the mismatch of the mirror between $I_{REF}$ and the output current, as the current ratio has a limited \mbox{1.5-$\%$} ($\sigma/\mu$) evaluated from 10$^3$ MC simulations. As hypothesized in \cite{Lefebvre_2022_ESSCIRC} and confirmed in this work, this reduction is linked to global process variations and more specifically to skewed process corners of the ULL and LVT I/O devices $M_8$ and $M_9$ constituting the body$\:$/$\:$back-gate generator, as illustrated in Figs.~\ref{fig:23_sim_meas_iref_cross_process}(a) and \ref{fig:24_sim_meas_iref_vs_T}(a). Details of the 20 dies are depicted in Fig.~\ref{fig:21_meas_iref_vs_vdd}(b), and histograms of $I_{REF}$ at 1.8~V and LS from 0.9 to 1.8~V are respectively presented in Figs.~\ref{fig:21_meas_iref_vs_vdd}(c) and (d). Fig.~\ref{fig:21_meas_iref_vs_vdd}(c) features a \mbox{9.20-$\%$} ($\sigma/\mu$) larger than the \mbox{6.66-$\%$} value obtained in Section~\ref{subsec:4C_final_implementation_performance}, which takes both mismatch and statistical process variations into account. Assuming the mean has been perfectly estimated, the \mbox{99-$\%$} confidence interval (CI) for the variance gives a CI for ($\sigma/\mu$) equal to [7.3~$\%$;$\:$15.3~$\%$], meaning that the limited sample size cannot explain the ($\sigma/\mu$) discrepancy. The lower $I_{REF}$ and higher ($\sigma/\mu$) are rather explained by a CWT offset $\Delta V_T$ lower than its nominal value [Figs.~\ref{fig:23_sim_meas_iref_cross_process}(a) and (c)], the larger variability being linked to an increased sensitivity $S_{I_{REF}}$ as $M_{1-2}$'s inversion level is reduced. Finally, Fig.~\ref{fig:21_meas_iref_vs_vdd}(d) reveals that the measured LS is around 0.39~$\%$/V, which is close to the 0.26~$\%$/V obtained in post-layout simulation.\\
\indent Furthermore, Fig.~\ref{fig:22_meas_iref_vs_T} gauges the temperature dependence of $I_{VDD}$ and $I_{REF}$ at 1.8~V. Fig.~\ref{fig:22_meas_iref_vs_T}(a) highlights the second order behavior of $I_{REF}$ in simulation and measurement, as well as a slighly PTAT trend observed in measurement only. $I_{VDD}$ amounts to 8.8~nA in simulation (resp. 6.4~nA in measurement), leading to a \mbox{7.8-nW} (resp. \mbox{5.8-nW}) power consumption at 0.9~V. Details of the 20 dies are portrayed in absolute value in Fig.~\ref{fig:22_meas_iref_vs_T}(b), and normalized by the value of $I_{REF}$ at 25$^\circ$C in Fig.~\ref{fig:22_meas_iref_vs_T}(c). The measured TC is comprised between 440 and 760~ppm/$^\circ$C, with an average of 565~ppm/$^\circ$C [Fig.~\ref{fig:22_meas_iref_vs_T}(d)]. Skewed process variations of ULL and LVT I/O devices reliably explain the TC deterioration, as a decreased (resp. increased) $\Delta V_T$ compared to its nominal value results in a slightly PTAT (resp. CTAT) behavior [Figs.~\ref{fig:23_sim_meas_iref_cross_process}(b) and (c)]. Besides, Fig.~\ref{fig:24_sim_meas_iref_vs_T} depicts the process corners showing the closest agreement with the measurements in terms of $I_{REF}$ and TC, with TCs of 446 and 557~ppm/$^\circ$C observed in the (TT, SF, typ.) and (FS, TT, typ.) process corners, respectively [Fig.~\ref{fig:24_sim_meas_iref_vs_T}(a)]. These results much better coincide with the measured \mbox{565-ppm/$^\circ$C} TC than the nominal corner [Fig.~\ref{fig:24_sim_meas_iref_vs_T}(b)]. Finally, Fig.~\ref{fig:24_sim_meas_iref_vs_T} suggests that the measurements are better explained by a minimum process for the parasitic nwell/psub diode $D_{VX}$ than by a typical process.
\setlength{\tabcolsep}{2pt}
\begin{table*}[!t]
\centering
\caption{Comparison table of temperature-independent nA-range current references.}
\vspace{-0.1cm}
\label{table:soa_nanoamp_range}
\resizebox{\linewidth}{!}{%
\begin{threeparttable}
\begin{scriptsize}
\begin{tabular}[t]{lcccccccccccccccc}
\toprule
& Huang & Far & Kim & Cordova & Santamaria & Aminzadeh & Kayahan & Dong & Ji & Wang & Wang & Huang & Lee & \multicolumn{3}{c}{\textbf{Lefebvre}}\\
& \cite{Huang_2010} & \cite{Far_2015} & \cite{Kim_2016} & \cite{Cordova_2017} & \cite{Santamaria_2019} & \cite{Aminzadeh_2022} & \cite{Kayahan_2013} & \cite{Dong_2017} & \cite{Ji_2017} & \cite{Wang_2019_VLSI} & \cite{Wang_2019_TCAS} & \cite{Huang_2020} & \cite{Lee_2020} & \multicolumn{3}{c}{\textbf{This work}}\\
\midrule
Publication & ISCAS & ROPEC & ISCAS & ISCAS & ISCAS & AEU & TCAS-I & ESSCIRC & ISSCC & VLSI-DAT & TCAS-I & TCAS-II & JSSC & \multicolumn{3}{c}{JSSC}\\
Year & 2010 & 2015 & 2016 & 2017 & 2019 & 2022 & 2013 & 2017 & 2017 & 2019 & 2019 & 2020 & 2020 & \multicolumn{3}{c}{2022}\\
\cmidrule(l){15-17}
Type of work & Sim. & Sim. & Sim. & Sim. & Sim. & Sim. & Silicon & Silicon & Silicon & Silicon & Silicon & Silicon & Silicon & Sim. & Sim. & Silicon\\
Number of samples & N/A & N/A & N/A & N/A & N/A & N/A & 90 & 32\tnote{$\ast$} & 10 & 10 & 16 & 10 & 10 & N/A & N/A & 20\\
\midrule
Technology & 0.18$\mu$m & 0.18$\mu$m & 0.13$\mu$m & 0.18$\mu$m & 0.18$\mu$m & 0.18$\mu$m & 0.35$\mu$m & 0.18$\mu$m & 0.18$\mu$m & 0.18$\mu$m & 0.18$\mu$m & 0.18$\mu$m & 0.18$\mu$m & 65nm & \multicolumn{2}{c}{22nm FDSOI}\\
$I_{REF}$ [nA] & 2.05 & 14 & 27 & 10.86 & 2.66 & 6.7 & 25 & 35.02 & 6.64 & 6.46 & 9.77 & 11.6 & 1 & 1.10 & 1.25 & 0.9\\
Power [nW] & 5.1 & 150 & N/A & 30.5 & 26 & 51 & 28500 & 1.02 & 9.3 & 15.8 & 28 & 48.64 & 4.5/14\tnote{$\triangleright$} & 5.41 & 7.84 & 5.81\\
& $@$0.85V & $@$1V & & $@$0.9V & $@$2V & $@$1.1V & $@$5V & $@$1.5V & $@$N/A & $@$0.85V & $@$0.7V & $@$0.8V & $@$1.5V & $@$0.7V & \multicolumn{2}{c}{$@$0.9V}\\
Area [mm$^2$] & N/A & \textcolor{ECS-Blue}{\textbf{0.0102}} & N/A & \textcolor{ECS-Blue}{\textbf{0.01}} & \textcolor{ECS-Blue}{\textbf{0.0093}} &\textcolor{ECS-Red}{\textbf{0.46}} & \textcolor{ECS-Blue}{\textbf{0.0053}} & 0.0169 & \textcolor{ECS-Red}{\textbf{0.055}} & \textcolor{ECS-Red}{\textbf{0.062}} & \textcolor{ECS-Red}{\textbf{0.055}} & \textcolor{ECS-Red}{\textbf{0.054}} & \textcolor{ECS-Red}{\textbf{0.332}} & \textcolor{ECS-Blue}{\textbf{0.0021}} & \multicolumn{2}{c}{\textcolor{ECS-Blue}{\textbf{0.0132}}}\\
\midrule
Supply range [V] & 0.85 -- 2.2 & 1 -- 3.3 & 1.2 & 0.9 -- 1.8 & 2 -- 3.3\tnote{$\diamond$} & 1.1 -- 1.8 & 5 & 1.5 -- 2.5 & 1.3 -- 1.8 & 0.85 -- 2 & 0.7 -- 1.2 & 0.8 -- 2 & 1.5 -- 2 & 0.7 -- 1.2 & \multicolumn{2}{c}{0.9 -- 1.8}\\
LS [$\%$/V] & 1.35 & \textcolor{ECS-Blue}{\textbf{0.1}} & N/A & \textcolor{ECS-Blue}{\textbf{0.54}} & \textcolor{ECS-Red}{\textbf{11}}\tnote{$\diamond$} &\textcolor{ECS-Blue}{\textbf{0.03}} & \textcolor{ECS-Red}{\textbf{150}} & 3 & 1.16 & 4.15 & \textcolor{ECS-Blue}{\textbf{0.6}} & 1.08 & 1.4 & \textcolor{ECS-Blue}{\textbf{0.69}} & \textcolor{ECS-Blue}{\textbf{0.26}} & \textcolor{ECS-Blue}{\textbf{0.39}}\tnote{$\triangleleft$}\\
\midrule
Temperature range [$^\circ$C] & 0 -- 150 & 0 -- 70 & -30 -- 150 & -20 -- 120 & -40 -- 125 & -40 -- 120 & 0 -- 80 & -40 -- 120 & 0 -- 110 & -10 -- 100 & -40 -- 125 & -40 -- 120 & -20 -- 80 & -40 -- 85 & \multicolumn{2}{c}{-40 -- 85}\\
TC [ppm/$^\circ$C] & \textcolor{ECS-Blue}{\textbf{91}} & \textcolor{ECS-Blue}{\textbf{20}} & 327 & \textcolor{ECS-Blue}{\textbf{108}} & \textcolor{ECS-Blue}{\textbf{182}} & \textcolor{ECS-Blue}{\textbf{40.33}} & \textcolor{ECS-Blue}{\textbf{128}}/250\tnote{$\star$} & 282 & 680/283\tnote{$\dagger$} & \textcolor{ECS-Blue}{\textbf{138}}\tnote{$\ddagger$} & \textcolor{ECS-Blue}{\textbf{149.8}} & \textcolor{ECS-Blue}{\textbf{169}} & 289/265\tnote{$\triangleright$} & \textcolor{ECS-Blue}{\textbf{213}} & \textcolor{ECS-Blue}{\textbf{203}} & 565\tnote{$\triangleleft$}\\
\midrule
$I_{REF}$ var. (process) [$\%$] & 7.5 & N/A & 3.7 & 15.8/11.6\tnote{$\dagger$} & N/A & N/A & 8/1.22\tnote{$\star$} & 4.7 & N/A & N/A & +11.7/-8.7\tnote{$\diamond$} & +17.6/-10.3\tnote{$\triangleleft$} & N/A & +19.1/-7.2 & +9.9/-9.5 & N/A\\
$I_{REF}$ var. (mismatch) [$\%$] & N/A & \textcolor{ECS-Red}{\textbf{5.8}} & N/A & N/A & \textcolor{ECS-Red}{\textbf{20.30}} & \textcolor{ECS-Blue}{\textbf{0.70}} & \textcolor{ECS-Blue}{\textbf{1.4}} (sim.) & \textcolor{ECS-Blue}{\textbf{1.6}} & 4.07/\textcolor{ECS-Blue}{\textbf{1.19}}\tnote{$\star$} & 3.33 & \textcolor{ECS-Blue}{\textbf{1.6}} & 4.3 & \textcolor{ECS-Blue}{\textbf{1.26}}/\textcolor{ECS-Blue}{\textbf{0.25}}\tnote{$\dagger$} & 2.92 & \textcolor{ECS-Red}{\textbf{6.39}} & \textcolor{ECS-Red}{\textbf{9.20}}\\
\midrule
Trimming & No & No & No & Yes$\:$(6b) & No & No & No & No & Yes & No & Yes$\:$(5b) & Yes$\:$(6b) & Yes$\:$(27b) & No & No & No\\
Special components & No & No & Res. & \textcolor{ECS-Red}{\textbf{ZVT}} & No & Res., \textcolor{ECS-Red}{\textbf{BJT}} & No & Res., \textcolor{ECS-Red}{\textbf{BJT}} & No & Res. & Res., & Res. & No & No & No & No\\
\bottomrule
\end{tabular}
\end{scriptsize}
\begin{footnotesize}
\begin{tablenotes}
	\item[$\ast$] 16 dies for the TT process corner and 4 dies for each of the FF, SS, SF and FS process corners.
	\item[$\star$] Simulated and measured values.
	\item[$\dagger$] Before and after trimming.
	\item[$\ddagger$] Best TC, the average one is not reported.
	\item[$\diamond$] Estimated from figures.
	\item[$\triangleleft$] Mean measured value across the 20 dies.
	\item[$\triangleright$] For 25 and 2.5 minutes between two calibrations.
\end{tablenotes}
\end{footnotesize}
\end{threeparttable}%
}
\end{table*}

\vspace{-0.25cm}
\section{Comparison to the State of the Art}
\label{sec:6_comparison_to_the_state_of_the_art}
\begin{figure}[!t]
	\centering
	\includegraphics[width=.45\textwidth]{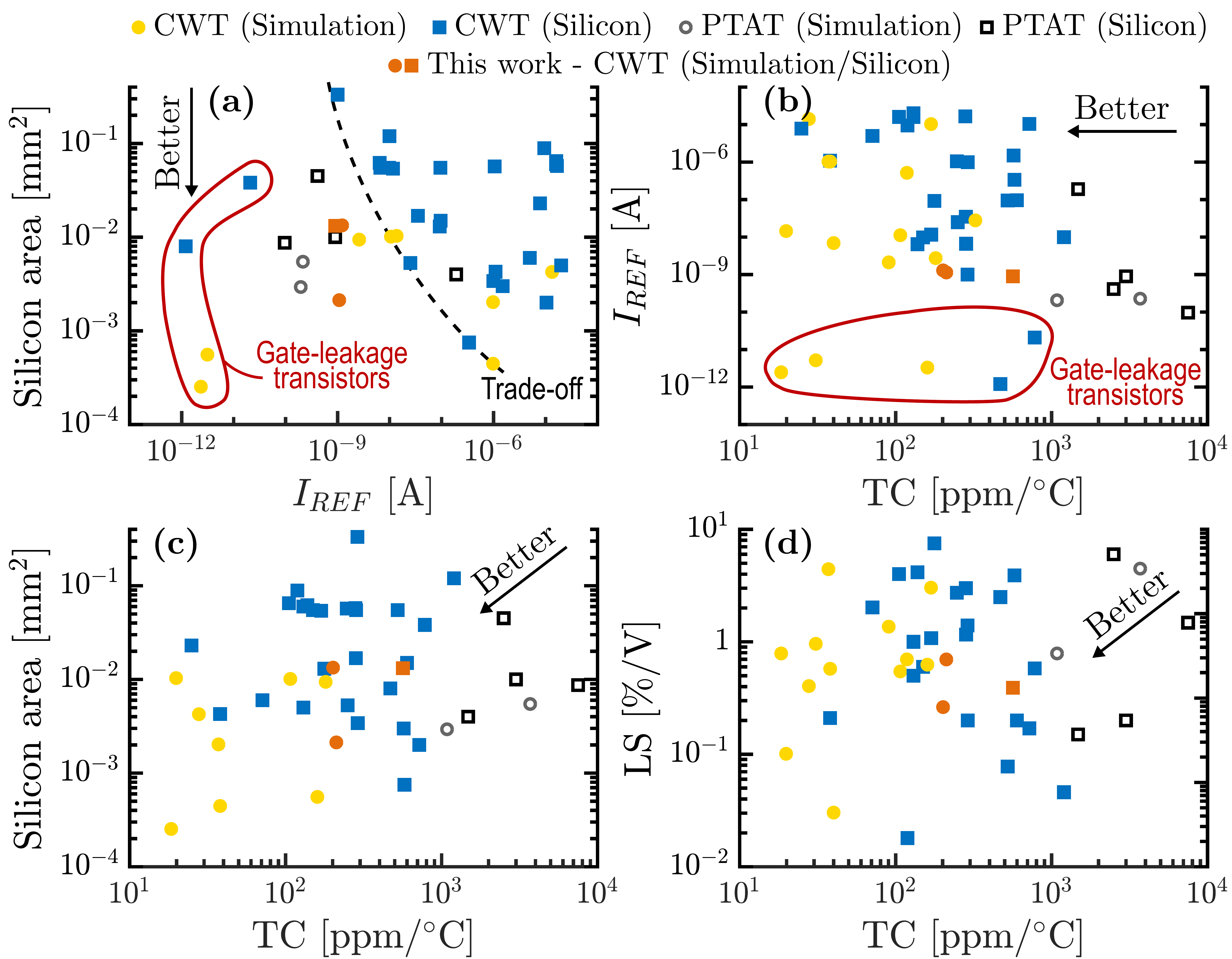}
	\caption{Trade-offs between (a) area and $I_{REF}$, (b) $I_{REF}$ and TC, (c) area and TC, and (d) LS and TC, based on the state of the art of current references.}
	\label{fig:25_comparison_to_soa}
\end{figure}

This section compares our work to the state of the art of current references through some important trade-offs, illustrated in Fig.~\ref{fig:25_comparison_to_soa}, and focuses on nA-range CWT references in Table~\ref{table:soa_nanoamp_range}. We report post-layout simulation results in \mbox{65-nm} bulk and \mbox{22-nm} FDSOI, and measurement results in \mbox{22-nm} FDSOI, coresponding to the three orange markers in Fig.~\ref{fig:25_comparison_to_soa}. Firstly, the 65-nm design surpasses all references in Table~\ref{table:soa_nanoamp_range}, simulated or fabricated, with a \mbox{0.0021-mm$^2$} area, and offers a 25$\times$ reduction compared to prior art of fabricated references. Then, the \mbox{0.0132-mm$^2$} silicon area in 22~nm outperforms all other fabricated references in Table~\ref{table:soa_nanoamp_range} by at least a factor 4$\times$, except for \cite{Kayahan_2013} whose \mbox{28.5-$\mu$W} power consumption is prohibitive in most applications.
Fig.~\ref{fig:25_comparison_to_soa}(a) further demonstrates that the proposed references are almost unchallenged in terms of silicon area among nA-range CWT references, and lie beyond the trade-off of conventional references. Only \cite{Santamaria_2019} is competitive in terms of silicon area with a \mbox{0.0093-mm$^2$ design}, as it relies on diode-connected and zero-$V_{GS}$ transistors as voltage-to-current converter. However, it suffers from a strong degradation of other FoMs with a 2-V $V_{DD,\mathrm{min}}$, an \mbox{11-$\%$/V} LS and a \mbox{20.3-$\%$} ($\sigma/\mu$). The low area overhead of the proposed reference is linked to the fact that it relies solely on transistors, contrary to other references which include area-hungry resistors. Besides, the implementation in advanced technology nodes further helps reducing the absolute area overhead. They reach the lowest reference current among CWT references at the exception of gate-leakage-based ones, which could be modified to reach a nA-range current but only at the cost of a $10^2$-to-$10^3\times$ larger area.\\
\indent In Table~\ref{table:soa_nanoamp_range}, power consumption and minimum supply voltage are on par with the state of the art, while LS is among the best for nA-range references with a sub-1-$\%$/V value obtained by low-voltage cascoding for both implementations, and backed by a large intrinsic gain $g_m/g_d$ in FDSOI \cite{Cathelin_2017}. Next, in terms of TC, the \mbox{213-} and \mbox{203-ppm/$^\circ$C} simulated values obtained for the proposed \mbox{65-} and \mbox{22-nm} designs are faintly larger than the TC of other fabricated references [Table~\ref{table:soa_nanoamp_range} and Fig.~\ref{fig:25_comparison_to_soa}(b)]. However, the design proposed in \cite{Wang_2019_TCAS, Huang_2010} relies on a temperature exponent of carrier mobility close to two and cannot be ported to most technologies, whereas \cite{Wang_2019_TCAS, Huang_2020, Ji_2017} are area-hungry due to the use of resistors and/or OTAs, and \cite{Lee_2020} due to the use of a current DAC and a low-TC reference. The TC degradation of the proposed 22-nm design to 565~ppm/$^\circ$C in measurement calls for a TC calibration mechanism, extensively used by fabricated references in Table~\ref{table:soa_nanoamp_range} \cite{Wang_2019_TCAS, Huang_2020, Ji_2017, Lee_2020}, to retrieve quasi simulation-level performance. Possible implementations could for example tune the current ratio between $M_5$ and $M_4$ using a binary-weighted mirror \cite{Huang_2020} to change $K_{PTAT}$, or the number of transistors connected in series to implement $M_1$ \cite{Zhuang_2020} to change $\alpha$. In Fig.~\ref{fig:25_comparison_to_soa}(c), the proposed references fare quite well against prior art but can simply not compete against gate-leakage-based (resp. resistor-based) references in the pA (resp. $\mu$A) range. Finally, Fig.~\ref{fig:25_comparison_to_soa}(d) demonstrates that the proposed designs are competitive with most existing references in terms of LS and TC. The main drawback of the proposed \mbox{22-nm} design is its large sensitivity to mismatch and process variations, with a \mbox{9.20-$\%$} ($\sigma/\mu$) due to a large $S_{I_{REF}}$, which could be improved by increasing the sizes of $M_{6-7}$, or by increasing $\Delta V_T$ as it is done in the \mbox{65-nm} design, whose simulated variability scales down to 2.92~$\%$.\looseness=-1

\vspace{-0.25cm}
\section{Conclusion}
\label{sec:7_conclusion}
In this work, we demonstrated a CWT current reference operating in the nA range. It relies on an SCM biased by a PTAT voltage with a CWT offset voltage, which stems from the $V_T$ difference between transistors of the same $V_T$ type, one of them being forward body-biased, through its body in bulk or back-gate in FDSOI. Then, we presented a comprehensive methodology for sizing the proposed reference based on the ACM model and validated it through post-layout simulations in \mbox{65-nm} bulk and \mbox{22-nm} FDSOI. The \mbox{65-nm} reference produces a \mbox{1.1-nA} current with an LS of 0.69~$\%$/V and a TC of 213~ppm/$^\circ$C, and power and area overheads of 5.4~nW and 0.0021~mm$^2$. Lastly, the \mbox{22-nm FDSOI} design was fabricated and generates a \mbox{0.9-nA} current with a \mbox{0.39-$\%$/V} LS and a {565-ppm/$^\circ$C} TC, while consuming 5.8~nW and occupying a silicon area of 0.0132~mm$^2$. The TC degradation from simulation to measurement, together with the large \mbox{9.20-$\%$} variability due to mismatch and process variations, ask for a TC calibration scheme to be explored in further work.


%

\vspace{-0.25cm}
\appendices
\section{Analytical Expressions of Voltage $V_X$}
\label{sec:8_appendix}
\indent Applying the ACM equations to transistors $M_{1-2}$ forming the SCM [Fig.~\ref{fig:2_basic_schematic}(b)] leads to two distinct equations. The first one is obtained by defining $\alpha \triangleq i_{f1}/i_{f2} > 1$ and expresses voltage $V_X$ as
\begin{IEEEeqnarray}{RCL}
	V_X & = & nU_T \Bigg[\left(\sqrt{1+\alpha i_{f2}}-\sqrt{1+i_{f2}}\right)\IEEEnonumber\\
	& & + \log\left(\frac{\sqrt{1+\alpha i_{f2}}-1}{\sqrt{1+i_{f2}}-1}\right)\Bigg]\textrm{,}\label{eq:vx_SCM_if1}
\end{IEEEeqnarray}
while the second one relates $\beta \triangleq i_{r1}/i_{f2} \in [0;1]$ to voltage $V_X$ and amounts to
\begin{IEEEeqnarray}{RCL}
	V_X(n-1) & = & nU_T \Bigg[\left(\sqrt{1+i_{f2}}-\sqrt{1+\beta i_{f2}}\right)\IEEEnonumber\\
	& & + \log\left(\frac{\sqrt{1+i_{f2}}-1}{\sqrt{1+\beta i_{f2}}-1}\right)\Bigg]\textrm{.}\label{eq:vx_SCM_ir1}
\end{IEEEeqnarray}


\vspace{-0.25cm}
\section*{Acknowledgments}
The authors would like to thank Pierre G\'{e}rard for the measurement testbench, El\'{e}onore Masarweh for the microphotograph, and ECS group members for their proofreading.

\ifCLASSOPTIONcaptionsoff
  \newpage
\fi



%
\bibliographystyle{IEEEtran}
\bibliography{Lefebvre_JSSC_2022_ICare}
\vspace{-1cm}

%

\begin{IEEEbiography}[{\includegraphics[width=1in,height=1.25in,clip,keepaspectratio]{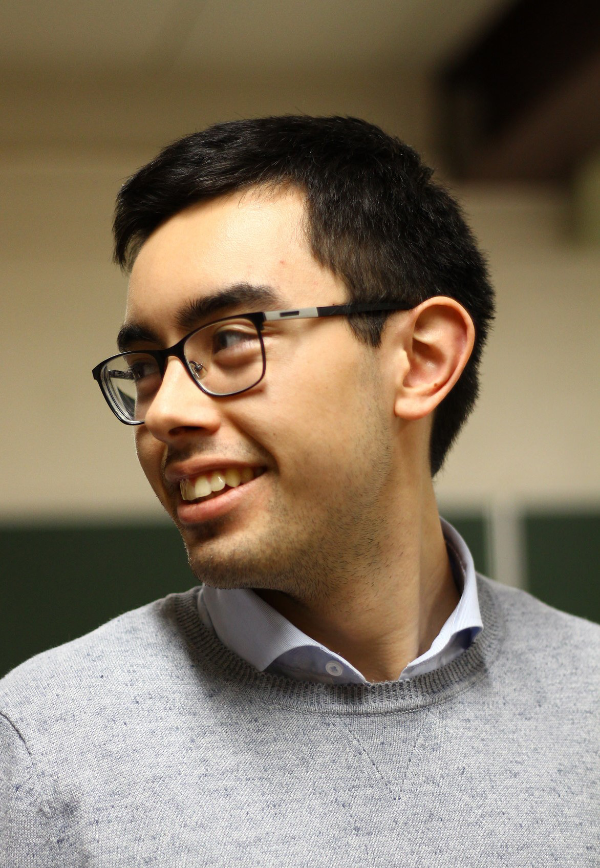}}]{Martin Lefebvre} (Graduate Student Member, IEEE) received the M.Sc. degree (summa cum laude) in Electromechanical Engineering from the Universit\'e catholique de Louvain (UCLouvain), Louvain-la-Neuve, Belgium, in 2017, where he is currently pursuing the Ph.D. degree, under the supervision of Prof. D. Bol. His current research interests include hardware-aware machine learning algorithms, low-power mixed-signal vision chips for embedded image processing, and ultra-low-power current reference architectures. He serves as a reviewer for various conferences and journals, including  IEEE Trans. on Biomed. Circuits and Syst., IEEE Trans. on VLSI Syst., Int. Symp. on Circuits and Syst. and Asia Pacific Conf. on Circuits and Syst.
\end{IEEEbiography}
\vspace{-1cm}

\begin{IEEEbiography}[{\includegraphics[width=1in,height=1.25in,clip,keepaspectratio]{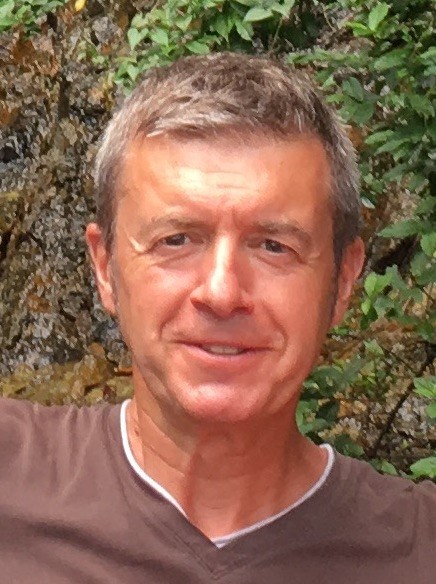}}]{Denis Flandre}
(Senior Member, IEEE) received the M.Sc. degree in Electrical Engineering, the Ph.D. degree and the Research Habilitation, from UCLouvain, Louvain-la-Neuve, Belgium, in 1986, 1990 and 1999, respectively. His doctoral research was on the modeling of silicon-on-insulator (SOI) MOS devices for characterization and circuit simulation, his post-doctoral thesis on a systematic and automated synthesis methodology for MOS analog circuits. Since 2001, he is full-time Professor at UCLouvain. He is involved in the research and development of SOI MOS devices, digital and analog circuits, as well as sensors, MEMS and solar cells, for special applications, more specifically ultra low-voltage low-power, microwave, biomedical, radiation-hardened and high-temperature electronics and microsystems. He has authored or co-authored more than 1000 technical papers or conference contributions. He is co-inventor of 12 patents. He has organized or lectured many short courses on SOI technology, devices and circuits in universities, industrial companies and conferences. He has received several scientific prizes and best paper awards. He has participated or coordinated numerous research projects funded by regional and European institutions. He has been a member of several EU Networks of Excellence on high-temperature electronics, SOI technology, nanoelectronics and micro-nano-technology. Prof. Flandre is a co-founder of CISSOID, a spin-off company of UCLouvain focusing on SOI and high-reliability integrated circuit design and products. He is scientific advisor of 3 other start-ups : INCIZE (semiconductor characterization and modeling for design of digital, analog/RF and harsh environment applications), e-peas (energy harvesting and processing solutions for longer battery life, increased robustness in all IoT applications) and VOCSens (smart gas sensing solutions from edge to cloud).
\end{IEEEbiography}
\vspace{-1cm}

\begin{IEEEbiography}[{\includegraphics[width=1in,height=1.25in,clip,keepaspectratio]{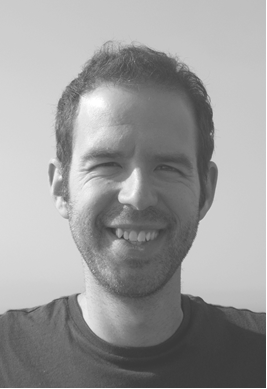}}]{David Bol} (Senior Member, IEEE) is an Associate Professor at UCLouvain. He received the Ph.D. degree in Engineering Science from UCLouvain in 2008 in the field of ultra-low-power digital nanoelectronics. In 2005, he was a visiting Ph.D. student at the CNM, Sevilla, and in 2009, a post-doctoral researcher at intoPIX, Louvain-la-Neuve. In 2010, he was a visiting post-doctoral researcher at the UC Berkeley Lab for Manufacturing and Sustainability, Berkeley. In 2015, he participated to the creation of e-peas semiconductors spin-off company. Prof. Bol leads the Electronic Circuits and Systems (ECS) group focused on ultra-low-power design of integrated circuits for environmental and biomedical IoT applications including computing, power management, sensing and wireless communications. He is actively engaged in a social-ecological transition in the field of ICT research with a post-growth approach. Prof. Bol has authored more than 150 papers and conference contributions and holds three delivered patents. He (co-)received four Best Paper/Poster/Design Awards in IEEE conferences (ICCD 2008, SOI Conf. 2008, FTFC 2014, ISCAS 2020) and supervised the Ph.D. thesis of Charlotte Frenkel who received the 2021 Nokia Bell Scientific Award and the 2021 IBM Innovation Award for her Ph.D. He serves as a reviewer for various IEEE journals and conferences and presented several keynotes in international conferences. On the private side, Prof. Bol pioneered the parental leave for male professors in his faculty, to spend time connecting to nature with his family. 
\end{IEEEbiography}




\end{document}